\documentclass[aps,prb,superscriptaddress,amsmath,amssymb,twocolumn,amsfonts,floatfix, longbibliography]{revtex4-2}
\usepackage{mathtools}
\usepackage{braket}
\usepackage[dvipsnames]{xcolor}
\usepackage{float}
\usepackage{subfigure}
\usepackage{upgreek}
\usepackage{pgffor}
\usepackage{float}
\usepackage{silence}
\usepackage[colorlinks=true,linktoc=page,linkcolor=blue,citecolor=magenta,urlcolor=Fuchsia]{hyperref}
\usepackage[normalem]{ulem}
\usepackage{sidecap,tikz}
\usepackage{orcidlink}

\newcommand{\beq}{\begin{equation}}
\newcommand{\eeq}{\end{equation}}
\newcommand{\bea}{\begin{eqnarray}}
\newcommand{\eea}{\end{eqnarray}}

\newcommand{\eqn}[1] {Eq.~(\ref{#1})}
\newcommand{\fig}[1]{Fig.~\ref{#1}}
\newcommand{\sect}[1]{Sec.~\ref{#1}}

\newcommand{\vect}[1]{\boldsymbol{\mathrm{#1}}}
\mathchardef\mhyphen="2D 
\newcommand{\noin}{\noindent}

\newcommand{\ua}{{\uparrow }}
\newcommand{\da}{{\downarrow }}
\newcommand{\eps}{{\epsilon}}

\newcommand{\non}{\nonumber}

\begin{document}
\title{Field-free superconducting diode effect in two-dimensional Shiba lattices}	
\author{Sayak Bhowmik}\thanks{These authors contributed equally to this work.}
\affiliation{Institute of Physics, Sachivalaya Marg, Bhubaneswar, Orissa 751005, India}
\affiliation{Homi Bhabha National Institute, Training School Complex, Anushakti Nagar, Mumbai 400094, India}

\author{{Dibyendu Samanta}\,\orcidlink{0009-0004-3022-7633}}\thanks{These authors contributed equally to this work.}
\affiliation{Department of Physics, Indian Institute of Technology, Kanpur 208016, India}

\author{{Ashis K. Nandy}\,\orcidlink{0000-0002-9296-2535}}
\email{aknandy@niser.ac.in}
\affiliation{School of Physical Sciences, National Institute of Science Education Research,An OCC of Homi Bhabha National Institute, Jatni 752050, India}
	
\author{{Arijit Saha}\,\orcidlink{0000-0002-6689-1657}}
\email{arijit@iopb.res.in}
\affiliation{Institute of Physics, Sachivalaya Marg, Bhubaneswar, Orissa 751005, India}
\affiliation{Homi Bhabha National Institute, Training School Complex, Anushakti Nagar, Mumbai 400094, India}

\author{{Sudeep Kumar Ghosh}\,\orcidlink{0000-0002-3646-0629}}
\email{skghosh@iitk.ac.in}
\affiliation{Department of Physics, Indian Institute of Technology, Kanpur 208016, India}
	
\date{\today}

\begin{abstract}

The superconducting diode effect (SDE) refers to non-reciprocal transport, where current flows without resistance in one direction but becomes resistive in the opposite direction, but its typical reliance on magnetic field hinders scalability and device integration. In this article, we present a theoretical framework for realizing a field-free SDE based on a two-dimensional (2D) Shiba lattice featuring a conical spin texture. Using the real-space Bogoliubov–de Gennes (BdG) calculations, we illustrate that the conical spin configuration alone is sufficient to break the necessary inversion and time reversal symmetries, enabling nonreciprocal supercurrent flow without any external magnetic field, yielding diode efficiency exceeding $40 \%$. Furthermore, we find that the efficiency of such a diode effect becomes strongly dependent on the direction of current flow, revealing a pronounced angular dependence that can be tuned by varying the pitches of the spin texture along the two spatial lattice directions. Our findings offer a pathway toward scalable, field-free superconducting components for non-dissipative electronics and quantum technologies.

\end{abstract}
	
\maketitle

\section{Introduction}

Diodes are fundamental to modern electronics and are characterized by a pronounced asymmetry in their voltage-current relationship, enabling unidirectional current flow. A superconducting analog of a conventional diode emerges when both inversion and time-reversal symmetries are broken, causing the critical current that drives the transition to the normal state to become direction-dependent, i.e., $I_c(\hat{n}) \ne I_c(-\hat{n})$~\cite{nadeem_2023,Nagaosa_2024}. Although, early studies observe such asymmetries in low-symmetry superconductors~\cite{Broussard_1988,Jiang_1994,Papon_2008}, recent advances in engineered superlattices~\cite{Narita_2024,ando_2020,sundaresh_2023}, bulk materials/ thin films~\cite{Wakatsuki_2017,nadeem_2023,Yuki_2020,Schumann_2020} etc., have significantly revived interest in the realization of efficient superconducting diode devices. Notably, platforms such as twisted multilayer graphene~\cite{lin_2022,Jaime_2023} and transition metal dichalcogenides~\cite{bauriedl_2022,Yun_2023} have emerged as promising candidates for harnessing and enhancing the efficient SDE due to their tunable band structures and strong spin–orbit interactions. Alongside these experimental efforts, a wide range of theoretical studies have proposed diverse mechanism for the realizations of SDE, including both tunnel-junction-based setups~\cite{Zhang_2022,Kokkeler_2022,Tanaka_2022,Legg_2023,Cuozzo_2024,Souto_2022,Cheng_2023,Steiner_2023,Costa_2023,Wei_2022} and junction-free superconducting phases~\cite{Daido_2022,Daido_2022_intrinsic,Yuan_2022,He_2022,Ili_2022,Scammell_2022,Zinkl_2022,He_2023,Zhai_2022,Jiang_2022,Picoli_2023,Legg_2022,Sayan_2024,bhowmik_2025,dibyendu_2025}, using approaches such as phenomenological Ginzburg–Landau theory and mean-field analysis. Crucially, the realization of SDE requires simultaneous breaking of  inversion and time-reversal symmetries~\cite{Wakatsuki_2017,Daido_2022_intrinsic,Nagaosa_2024}, which is typically accomplished through external magnetic fields~\cite{ando_2020,sundaresh_2023,Hou_2023,gupta_2023,Abhishek_2023} or magnetic proximity effects~\cite{Yun_2023,narita_2022,gutfreund_2023}. Recently, field-free SDE has also been proposed in rhombohedral graphene multilayers~\cite{chen_2025}, pointing toward viable strategies for practical and scalable superconducting diodes.

On the other hand, helical Shiba chains-engineered by depositing magnetic adatoms on top of $s$-wave superconducting substrates—have recently emerged as a compelling platform for realizing topological superconductivity hosting Majorana modes~\cite{Choy_2011,Nadj-Perge_2013,Pientka_2013,Vazifeh_2013,Heimes_2014}. The interplay between the helical magnetic order and superconducting pairing lead to the formation of localized Yu–Shiba–Rusinov (YSR) states, which hybridize into emergent Shiba bands residing within the superconducting gap. These bands can host robust topological superconducting phases characterized by Majorana zero modes localized at the ends of the one-dimensional (1D) chain. Notably, the SDE has recently been demonstrated in such helical Shiba chains under an applied magnetic field in 1D~\cite{Sayak_2025}. This phenomenon is attributed to the emergence of a finite-momentum Fulde–Ferrell–Larkin–Ovchinnikov (FFLO) superconducting state~\cite{Fulde_1964,larkin_1965}, driven by the combined effects of the helical spin texture and the applied magnetic field.

In junction-free systems, the SDE has typically been observed only under the influence of an external magnetic field,  limiting its suitability for practical device integration. To realize the field-free SDE, we consider a 2D conical spin texture deposited on the surface of a 3D $s$-wave superconductor. This heterostructure intrinsically breaks both time-reversal and inversion symmetries, key requirements for realizing the SDE, without relying on any applied magnetic field. Using a self-consistent BdG mean-field approach, we demonstrate that the system hosts a Fulde–Ferrell–Larkin–Ovchinnikov (FFLO) superconducting ground state. Within this phase, we uncover highly efficient diode behavior, with efficiencies exceeding $\sim 40~\%$ through suitable tuning of system parameters. Our proposal can be tested in existing hybrid magnetic–superconducting platforms where conical spin textures can be engineered~\cite{Weis}. Our findings establish a realistic and tunable platform for realizing field-free superconducting diodes in hybrid magnetic-superconducting heterostructures, opening promising avenues for experimental implementation.

The structure of the paper is as follows. In \sect{sec2}, we introduce the lattice Hamiltonian describing a 2D Shiba lattice with a conical spin texture. To gain analytical insight into the system, we derive both the low-energy continuum model and its lattice-regularized counterpart for our setup in \sect{sec3}. In \sect{sec4}, we analyze the emergence and key features of the FFLO superconducting state, followed by our main results on nonreciprocal charge transport in \sect{sec5}.  Finally, in Sec.~\ref{sec6}, we summarize our main results and discuss candidate material platforms for experimental realization, and an outlook for future directions.


\begin{figure}[t]
\centering 
\includegraphics[width=\columnwidth]{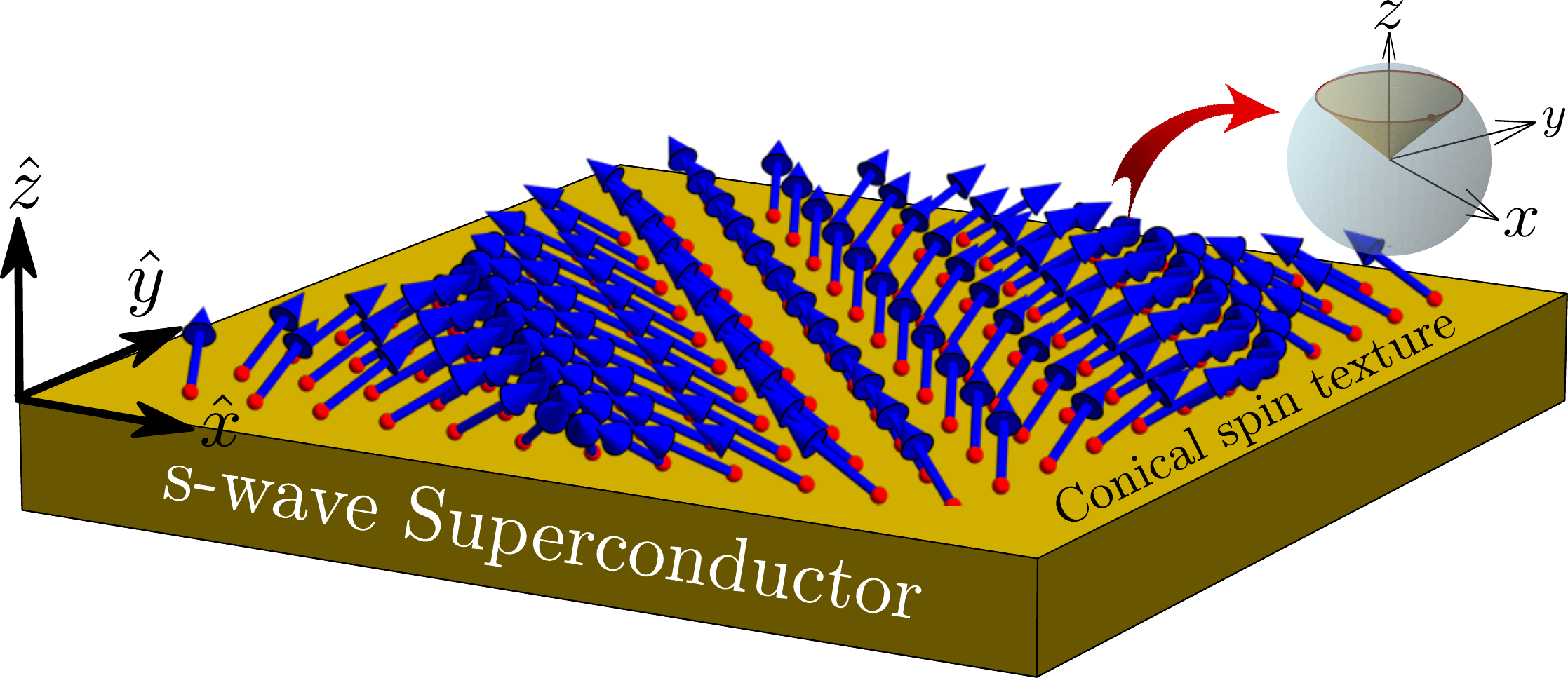}
\caption{\textbf{Schematic of a 2D Shiba lattice with conical spin texture:} Magnetic adatoms with spatially varying classical spins are placed on the surface of a three-dimensional (3D) $s$-wave superconductor, forming a conical spin texture. The orientation of the classical spins are highlighted in a Bloch sphere.}   
\label{fig:model}
\end{figure}



\section{Model Hamiltonian}\label{sec2}

We consider a 2D array of classical spins forming a conical texture, placed atop a 3D $s$-wave superconductor, as illustrated schematically in \fig{fig:model}. Our aim is to explore the superconducting ground state that arises in the resulting Shiba lattice from the interplay between classical magnetic impurities and $s$-wave pairing correlation. In the presence of a finite supercurrent, this system can host a FFLO superconducting state. To understand this physics,we model the parent $s$-wave superconductivity using an attractive on-site Hubbard interaction: $\mathcal{H}_I = - \frac{U}{2} \sum_{s,s^\prime} \int \, d^3\mathbf{r}\,c^{\dagger}_s(\mathbf{r}) c^{\dagger}_{s^\prime}(\mathbf{r}) c_{s^\prime}(\mathbf{r}) c_{s}(\mathbf{r})$, with interaction strength $U >0$ and $c_s(\mathbf{r})$ annihilates an electron with spin $s$ at position $\mathbf{r}$. Within the mean-field approximation, we decouple $\mathcal{H}_I$ in the $s$-wave FFLO channel  and assume that superconducting correlations are proximity induced in the spin texture. This approach allows us to treat the 2D spin texture as an effective superconducting system with a self-consistently determined pairing amplitude. In this framework, the 2D mean-field BdG Hamiltonian can be expressed in the Nambu basis $\Psi(\vect{r})$=$\begin{pmatrix} c_{\vect{r_i},\uparrow},\! & \! c_{\vect{r_i},\downarrow},\! &\! c^{\dagger}_{\vect{r_i},\downarrow},\! &\!\! -c^{\dagger}_{\vect{r_i},\uparrow} \end{pmatrix}^{\rm T}\!$ as follows:

\begin{equation}
\begin{aligned}
&H_L=\sum_{<i,j> ,\alpha}(t c^{\dagger}_{i,\alpha}c_{j,\alpha}+h.c.)+J\sum_{i,\alpha,\beta}(\vect{S}_{i}.\sigma)_{\alpha,\beta}c^{\dagger}_{i,\alpha}c_{i,\beta}  \\
&-\mu\sum_{i,\alpha}c^{\dagger}_{i,\alpha}c_{i,\alpha}+\sum_i(\Delta e^{i\vect{q}.\vect{r}_i}c^{\dagger}_{i,\uparrow}c^{\dagger}_{i,\downarrow}+h.c.)+\frac{\mathcal{A}}{U}|\Delta|^2.
\end{aligned}
\label{eq:lattice1}
\end{equation}

Here, $c_{i,\alpha}$ ($c^{\dagger}_{i,\alpha}$) represents the annihilation (creation) operators at site $i$ with position $r_i$ and spin $\alpha \in \uparrow, \downarrow$ and $\rm T$ denotes the transpose conjugation. The local spin vector $\vect{S}(\vect{r_i})$ at each site is considered as $\vect{S}(\vect{r}_i)$=$\lvert \vect{S_i} \rvert\begin{pmatrix} \sin[{\theta}]\cos[\phi(\vect{r_i})],\!&\! \sin[{\theta}]\sin[\phi(\vect{r_i})],\! &\! \cos[{\theta}] \end{pmatrix}$ where, $\phi(\vect{r_i})$=$\vect{g}\!\cdot \!\vect{r}_i$=$\left( g_x x_i+g_y y_i \right)$, represents the spatial variation of the spins with $\vect{g}$ being the pitch of the spin texture and $0< \theta < \pi/2$ ensures the cone phase of the corresponding spin texture. Here, $(i,j)$ represents the site index and further for simplicity we consider $|S_i|=1$ throughout the article. The $s$-wave FFLO superconducting order parameter is represented by: $\Delta e^{i\vect{q}.\vect{r_i}}$ where $\vect{q}$ denotes the Cooper-pair momentum. Here, $\mathcal{A}$ being the system area and $J$, and $t$ represent the exchange interaction strength, and hopping amplitude respectively. 

\section{Continuum model and effective lattice regularized Hamiltonian}\label{sec3}
To gain analytical insight into our numerical result of the lattice, we employ a low-energy description of the Hamiltonian Eq.~\eqref{eq:lattice1} 
in the continuum limit that takes the form:   

\begin{equation}
\begin{aligned}
    H_C &= \int d\mathbf{r} \, \Psi^\dagger(\mathbf{r}) \, \mathcal{H}(\mathbf{r}) \, \Psi(\mathbf{r}) + \frac{\mathcal{A}}{U}|\Delta|^2, \\
    \mathcal{H}(\mathbf{r}) &= -\frac{1}{2}( \nabla^2-\mu) \tau_z + J \, \mathbf{S}(\mathbf{r}) \cdot \boldsymbol{\sigma} + \Delta e^{i\vect{q}.\vect{r}} \, \tau_x \ .
\end{aligned}
\end{equation}

\noindent For simplicity, we set $\hbar = m = 1$ throughout our analysis. Here, $\pmb{\sigma}$ and $\pmb{\tau}$ are Pauli matrices acting in spin and particle-hole subspaces, respectively with other parameters are as described for Hamiltonian in Eq.~\eqref{eq:lattice1}. Furthermore, upon performing a local unitary transformation $U=e^{-\frac{i}{2} \phi(\mathbf{r}) \sigma_z}$~\cite{Richard_2022}, and followed by Fourier transformation, one can obtain the corresponding continuum momentum-space Hamiltonian: $H_C= \frac{1}{2}\sum_{\mathbf{k}} \psi^\dagger(\mathbf{k},\mathbf{q}) \mathcal{H}(\mathbf{k}, \mathbf{q}) \psi(\mathbf{k},\mathbf{q}) + \frac{A}{U}|\Delta|^2$, written in the  following BDG basis:  $\psi(\mathbf{k},\mathbf{q})=\left( c_{\mathbf{k}+ \mathbf{q}/2,\ua}, c_{\mathbf{k}+ \mathbf{q}/2,\da}, c_{-\mathbf{k}+ \mathbf{q}/2,\da}^\dagger, -c_{-\mathbf{k}+ \mathbf{q}/2,\ua}^\dagger \right)^{\rm T}$, where  

\begin{align}
    &\mathcal{H}(\mathbf{k}, \mathbf{q}) =
    \begin{bmatrix}
        h_{\mathbf{k}+\mathbf{q}/2} & \Delta \\
        \Delta & -h^*(-\mathbf{k} + \mathbf{q}/2)
    \end{bmatrix}, \nonumber \\
    &h_{\mathbf{k}} = \epsilon_{\mathbf{k},\tilde{\mathbf{g}}} + t~(\mathbf{g} \cdot \mathbf{k}) \sigma_z + J \sin(\theta) \sigma_x + J \cos(\theta) \sigma_z \ .
    \label{eq:k_space_hamiltonian}
\end{align}
\noindent Here, $\eps_{\mathbf{k},\mathbf{g}}= t~(\mathbf{k}^2 + \mathbf{\tilde{g}}^2) - \mu$, with $\mathbf{\tilde{g}}=\mathbf{g}/2$, where $t = \frac{\hbar^2}{2m}$ to ensure direct correspondence with the continuum limit. The spin texture in cone phase interestingly give rise to an effective  spin-orbit coupling (SOC) along with Zeeman terms, thus breaking both the time reversal and inversion symmetries, as evident from  Eq.~\eqref{eq:k_space_hamiltonian}. Unlike conventional SOC, its origin is nontrivial, stemming from the spatial variation of the spintexture rather than intrinsic relativistic effects. The Bogoliubov quasiparticle spectrum exhibits asymmetry intrinsically in the presence of a conical spin texture, whereas for the helical spin texture, an external magnetic field is required to induce spectral asymmetry (see Supplementary Material (SM) for details). These insights are crucial for understanding the mechanisms underlying the SDE.




The effective lattice regularized version of the continuum Hamiltonian can be obtained by applying the substitutions $k \rightarrow \sin(k)$ 
and $(1 - \frac{k^2}{2}) \rightarrow \cos(k)$ followed by an inverse Fourier transformation performed in the Hamiltonian described by Eq.~\eqref{eq:k_space_hamiltonian} leading to:  
\begin{align}
    H^{\prime} =& - \sum_{\substack{\mathbf{n} \\ \sigma, \sigma^\prime}} \sum_{\nu = x, y} c_{\sigma^\prime, \mathbf{n}+\hat{\nu}}^\dagger \bigg[ t\, \delta_{\sigma,\sigma^\prime} + \frac{i}{2}\tilde{g}_\nu (\sigma^z)_{\sigma'\sigma} + \text{h.c.}\bigg] c_{\sigma, \mathbf{n}} \nonumber \\
    &+\sum_{\sigma, \sigma^\prime, \mathbf{n}} c_{\sigma^\prime, \mathbf{n}}^\dagger \bigg[ \left( 4t + \frac{\tilde{g}_{x}^2 + \tilde{g}_{y}^2}{2} - \mu \right)\delta_{\sigma,\sigma^\prime} \nonumber \\
    &+ J \cos(\theta) (\sigma^z)_{\sigma^\prime \sigma} + J \sin(\theta) (\sigma^x)_{\sigma^\prime \sigma} \bigg] c_{\sigma, \mathbf{n}} \nonumber \\
    & + \sum_{\mathbf{n}} \left( \Delta e^{i \mathbf{q}\cdot\mathbf{n}} c_{\mathbf{n},\ua}^\dagger c_{\mathbf{n},\da}^\dagger + h.c. \right) + \frac{L_x L_y}{U} |\Delta|^2\,, \non \\
    =& \sum_{m} \left(E_m (\mathbf{q}) \gamma_{m}^\dagger \gamma_m - \frac{E_m}{2} \right) + \frac{L_x L_y}{U} |\Delta(\mathbf{q})|^2\;.
    \label{eq:ham_real}
\end{align}
\begin{figure*}[t]
\centering 
\includegraphics[width=2.07\columnwidth]{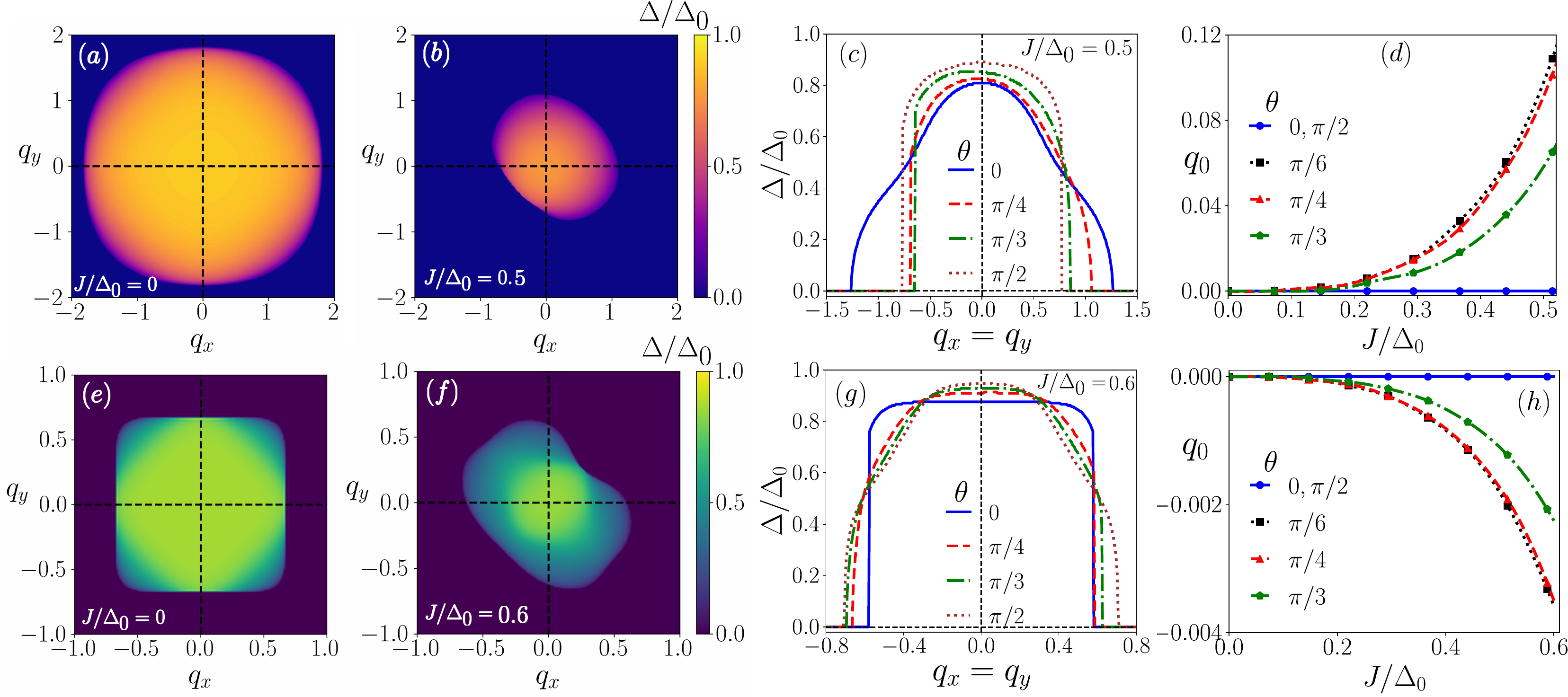}
\caption{\textbf{FFLO ground state for Shiba lattice with conical spin texture:} The self-consistent superconducting gap $\Delta$ is shown as a function of Cooper pair momentum $\mathbf{q}$ in the $q_x$–$q_y$ plane for $J/\Delta_0 = 0$ and $J/\Delta_0 = 0.5$ in (a) and (b), respectively, using the original lattice Hamiltonian [\eqn{eq:lattice1}] for $t/\Delta_0=0.5$ and $U/\Delta_0=2.56$. The variation of $\Delta$ along the $q_x = q_y$ direction for $J/\Delta_0 = 0.5$ and the ground-state momentum $q_0$ as a function of $J$ for different values of $\theta$ are shown in (c) and (d). The corresponding results based on the effective lattice regularized Hamiltonian [\eqn{eq:ham_real}] are presented in (e)–(h), with $t/\Delta_0 = 1.0$ and $U/\Delta_0 = 1.81$. The superconducting gap $\Delta$ in the $q_x$–$q_y$ plane for $\theta = \pi/4$ is plotted for $J/\Delta_0 = 0$ and $J/\Delta_0 = 0.6$ in (e) and (f), while (g) and (h) display $\Delta$ along $q_x = q_y$ and $q_0$ versus $J$ for various $\theta$. Other model parameters are taken as $(g_x,g_y,\mu/\Delta_0,\beta^{-1}/\Delta_0, \theta) = (\pi/2,\pi/2,1,0.1,\pi/4)$.}
\label{fig:self}
\end{figure*}

\noindent Here, $\hat{\nu}$ is the unit vector along the $\nu$-direction (with $\hat{x} = (1, 0)$ and $\hat{y} = (0, 1)$), and the system consists of $L_x \times L_y$ lattice sites. Also, $E_m(\mathbf{q})$ represents the energy of the $m^{\rm{th}}$ Bogoliubov quasiparticle state, while $\gamma_m^\dagger$ and $\gamma_m$ denote the corresponding quasiparticle creation and annihilation operators, respectively. Through self-consistent analysis, we demonstrate that the conical phase of the spintexture remarkably favours the emergence of FFLO superconduting state, as discussed in the subsequent sections.

\section{FFLO superconducting state}\label{sec4}
The condensation energy of the Shiba lattice, defined as the difference in free energy per unit area between the superconducting and normal states, is given by $\Omega(\mathbf{q}, \Delta) = F(\mathbf{q}, \Delta) - F(\mathbf{q}, 0)$, where, $F(\mathbf{q},\Delta)$ corresponds to the free energy density of the system given by

\begin{equation}
    F (\mathbf{q}, \Delta) = - \frac{1}{L_x L_y \beta} \sum_m \left[ 1 + e^{-\beta E_m(\mathbf{q})} \right] + \frac{|\Delta(\mathbf{q})|^2}{U}\ , \label{eq:free_energy}
\end{equation}

\noindent where, $\beta = (k_B T)^{-1}$ with $k_B$ being the Boltzmann constant and $T$ is the temperature.

For a given value of $\mathbf{q}$, the order parameter $\Delta(\mathbf{q})$ is determined by self-consistently solving the gap equation, which is derived by minimizing the condensation energy $\Omega(\mathbf{q}, \Delta)$ as defined in Eq.~(\ref{eq:free_energy}):
\begin{equation}
    \Delta(\mathbf{q}) = - \frac{U}{L} \sum_m \frac{\partial E_m}{\partial \Delta^*} n_F (E_m)\ . \label{eq:gap_equation}
\end{equation}

\noindent Here, $n_F(E_m) = \frac{1}{1 + e^{\beta E_m}}$ denotes the Fermi-Dirac distribution function of the Bogoliubov quasi-particles. 
To determine the true FFLO superconducting ground state, we self-consistently solve for $\Delta(\mathbf{q})$, and then optimize the condensation energy with respect to the Cooper pair momentum $\mathbf{q}$. The resulting optimal momentum, $\mathbf{q}_0$, characterizes the finite-momentum pairing in the FFLO phase. The parameter $\Delta_0$ represents the BCS gap of a 3D $s$-wave superconductor, and the Hubbard interaction strength $U$ is chosen such that the system remains in the weak-coupling BCS regime, with $\Delta_0 \sim 1$~meV.

The presence of spin texture in the conical phase ($J \ne 0,\ \theta \ne 0 \ne \pi/2$) inherently breaks both inversion and time-reversal symmetries as discussed before, thereby favoring an FFLO superconducting phase characterized by a finite Cooper pair momentum in the ground state. This is demonstrated in \fig{fig:self} via self-consistent mean-field analysis. The realization of such an FFLO phase is crucial for enabling non-reciprocal supercurrents, i.e., the superconducting diode effect. The self-consistent $\Delta(\mathbf{q})$ for a conical spin texture ($\theta = \pi/4$) is shown in Fig.~\ref{fig:self}, with panels (a),(b) corresponding to the original lattice Hamiltonian and (e),(f) to the lattice-regularized model. Notably, $\Delta(\mathbf{q})$ vanishes beyond a critical Cooper pair momentum $q_c(\delta)$, which depends on the polar angle $\delta$ in the $q_x$–$q_y$ plane. In the absence of exchange coupling ($J = 0$), the critical momentum is symmetric, i.e., $q_c(\delta) = q_c(\delta + \pi)$, as can be seen in \fig{fig:self}(a),(e). However, once the exchange coupling ($J$) is turned on, this symmetry is broken and $\Delta(\mathbf{q})$ becomes asymmetric, with $q_c(\delta) \ne q_c(\delta + \pi)$ [see \fig{fig:self}(b,f)]. Furthermore, to gain insights into our results, in \fig{fig:self}(c),(g) we display the variation of $\Delta(\mathbf{q})$ along the $q_x = q_y$ line for different spin textures ($\theta$ values). Note that, for planar ($\theta = \pi/2$) and trivial ($\theta = 0$) textures, the order parameter remains symmetric and the system does not support FFLO pairing. In contrast, for conical textures ($0 < \theta < \pi/2$), $\Delta(\mathbf{q})$ becomes asymmetric which is further illustrated in Fig.~\ref{fig:self}(d),(h) where the evolution of the Cooper pair momentum $q_0$ in the FFLO ground state is depicted as a function of exchange coupling $J$.
It is evident that, as $J$ increases, $q_0$ also enhances and exhibits a nonlinear growth beyond $J/\Delta_0 = 0.2$. Notably, for planar and trivial spin textures, $q_0$ remains essentially zero with increasing $J$, confirming the absence of FFLO pairing in these cases.

\section{Nonreciprocal transport}\label{sec5}
\begin{figure}[t]
\centering 
\includegraphics[width=1.03\columnwidth]{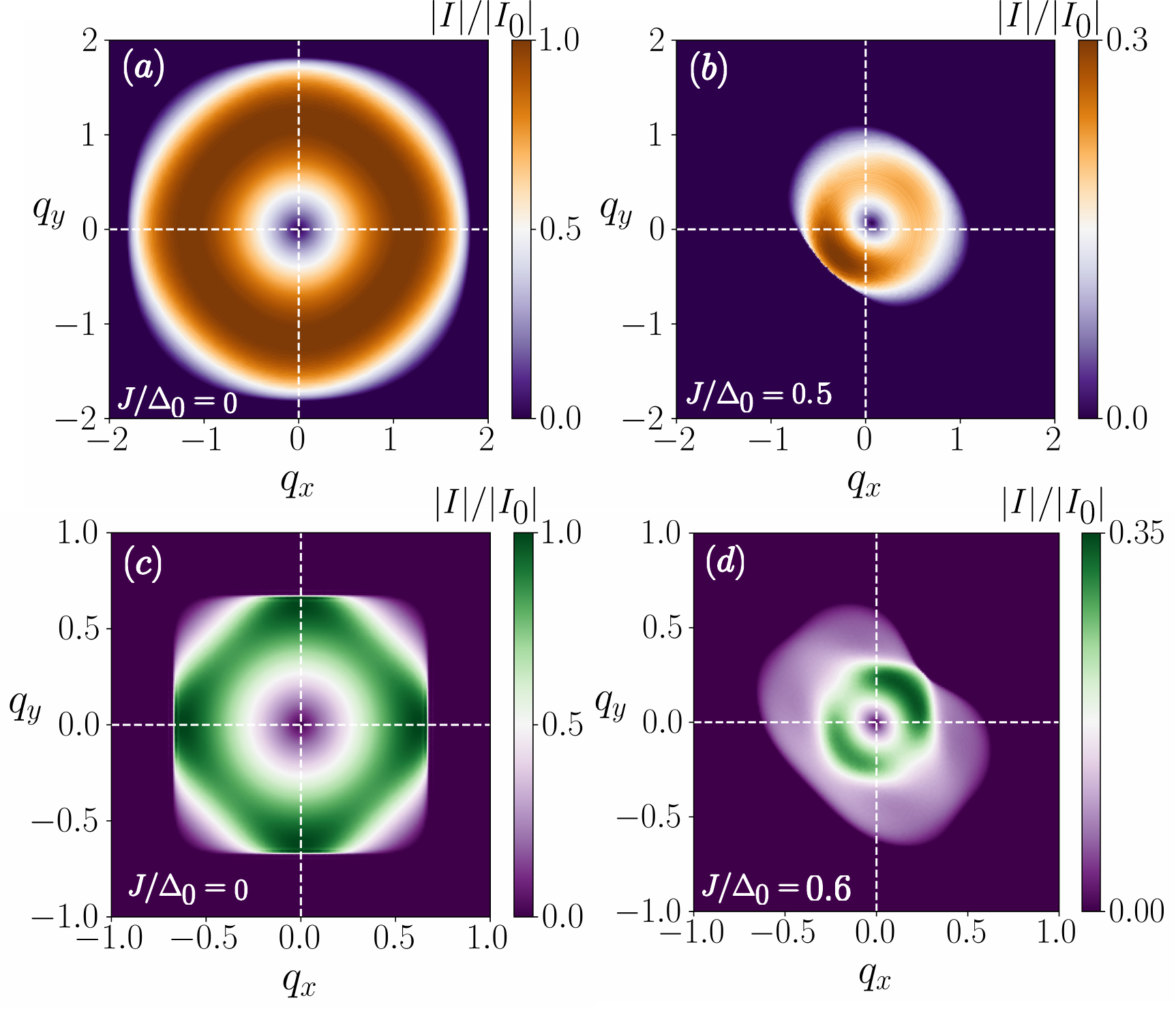}
\caption{\textbf{Super-current density profile for Shiba lattice with conical spin texture:} The super-current density $\vect{I}$ is shown in the $q_x$–$q_y$ plane for $J/\Delta_0 = 0$ and $0.5$ in (a) and (b), respectively, using the lattice Hamiltonian [\eqn{eq:lattice1}] for $t/\Delta_0 = 0.5$, and $U/\Delta_0 = 2.56$. Corresponding results from the lattice-regularized Hamiltonian [\eqn{eq:ham_real}] are presented in (c) and (d) for $J/\Delta_0 = 0$ and $0.6$, using $t/\Delta_0 = 1.0$ and $U/\Delta_0 = 1.81$. The current is normalized by $\vect{I}_0 = \vect{I}_c(J = 0, \mu/\Delta_0 = 1)$. Other parameters are $(g_x, g_y, \mu/\Delta_0,\theta) = (\pi/2, \pi/2, 1, \pi/4)$.}    
\label{fig:current}
\end{figure}

Having demonstrated the emergence of FFLO pairing in our proposed setup, we now turn our attention to the striking phenomenon of the SDE, a hallmark feature intrinsically linked to the FFLO ground state. In general, SDE is realized via the presence of unequal critical super-currents flowing in opposite directions: $I_c(\alpha)\ne I_c(\pi+\alpha)$, where $\alpha$ denotes the polar angle of the 2D super-current vector: $\vect{I}$. Here, ${I_c}$ corresponds to the maximum supercurrent the superconductor can sustain (de-pairing current) before making transition to the normal state.  Utilizing  the self-consistent solution of $\Delta(\vect{q})$, we then compute the super-current density following the bond current formalism, where the current flowing between sites $i$ and $i+\hat{\nu}$ can be defined as~\cite{Zhu_2016}:
\begin{figure}[h]
\centering 
\includegraphics[width=1.\columnwidth]{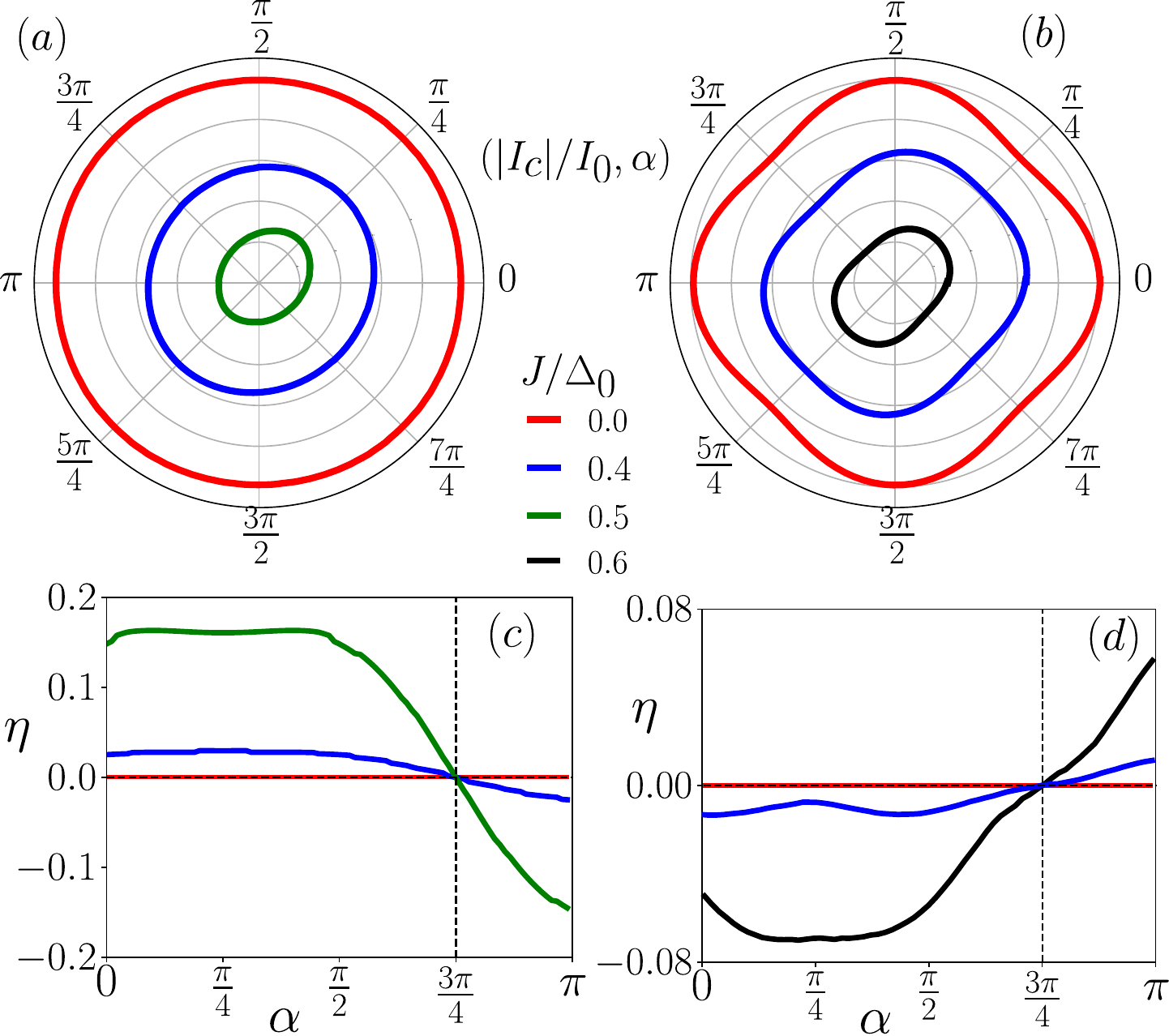}
\caption{\textbf{Non-reciprocal super-current and diode efficiency for Shiba lattice with conical spin texture:} The angular dependence ($\alpha$) of the critical current magnitude $|I_c|$ and diode efficiency $\eta$ is shown for various $J$ in (a) and (c) using the lattice Hamiltonian [\eqn{eq:lattice1}] with $t/\Delta_0 = 0.5$ and $U/\Delta_0 = 2.56$. Corresponding results based on the lattice-regularized Hamiltonian [\eqn{eq:ham_real}] with $t/\Delta_0 = 1.0$ and $U/\Delta_0 = 1.81$ are shown in (b) and (d). Other parameters are $(g_x, g_y, \mu/\Delta_0, \theta) = (\pi/2, \pi/2, 1, \pi/4)$.}
\label{fig:eff}
\end{figure}
\begin{align}
& I_{\mathbf{i}}^\nu = \frac{e}{i \hbar} \sum_{\sigma, \sigma'} \sum_m \Bigg\{ \Bigg[ 
- t'(\mathbf{i}, \sigma; \mathbf{i} + \hat{\nu}, \sigma') \, u_{\mathbf{i},\sigma}^{m*} \, u_{\mathbf{i} + \hat{\nu}, \sigma'}^m \, n_F(E_m) \nonumber \\
&\quad + t'^*(\mathbf{i}, \sigma; \mathbf{i} + \hat{\nu}, \sigma') \, v_{\mathbf{i},\sigma}^m \, v_{\mathbf{i} + \hat{\nu}, \sigma'}^{m*} \, (1 - n_F(E_m)) \Bigg] - \text{c.c.} \Bigg\}\ .
\end{align}

\begin{figure*}[t]
\centering 
\includegraphics[width=2\columnwidth]{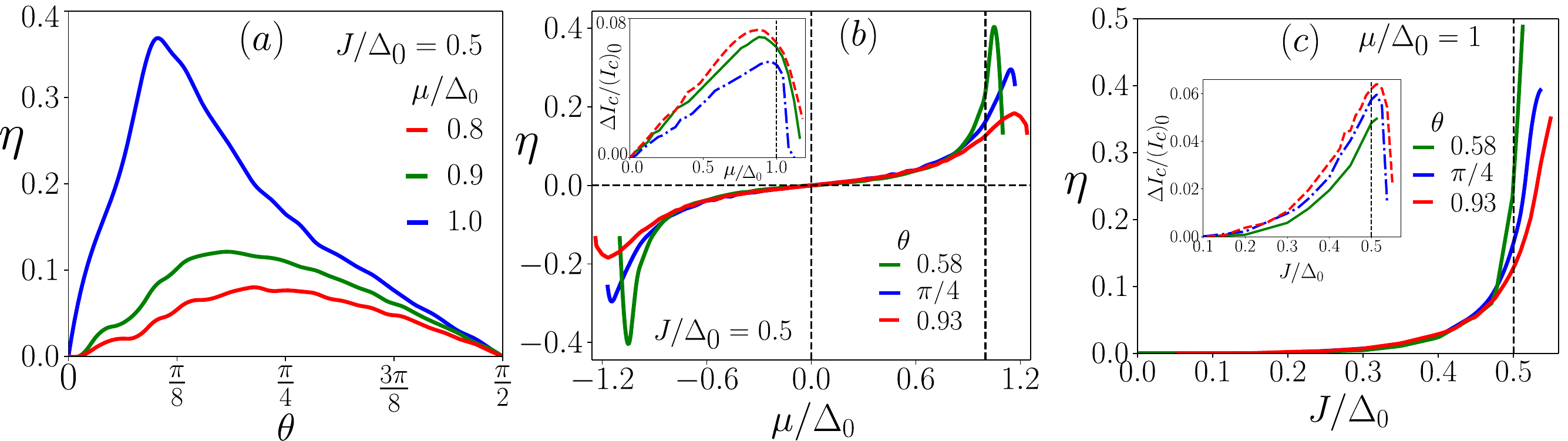}
\caption{\textbf{Optimizing diode efficiency:} The optimal diode efficiency is depicted as a function of $\theta$ for different values of $\mu$ with $J/\Delta_0=0.5$ in panel (a). The dependence of $\eta$ is displayed as a function of $\mu$ in panel (b) for the optimal choices of $\theta$ with $J/\Delta_0=0.5$. Panel (c) demonstrates $\eta$ as a function of $J$ for optimal choices of $\theta$ and $\mu/\Delta_0=1$. The measure of non-reciprocity (asymmetry) of critical currents $\Delta I_c=I_c(\alpha)-I_c(\alpha+\pi)$ is shown as a function of $\mu$ and $J$ in the insets of panels (b) and (c), respectively. Here, $(I_c)_0= I_c (J,\mu=0) $.  Other model parameters of the lattice Hamiltonian [Eq.~(\ref{eq:lattice1})] are chosen as $(g_x,g_y,\mu/\Delta_0,\theta,t/\Delta_0,\alpha)=(\pi/2,\pi/2,1,\pi/4,0.5,\pi/4)$.}  
\label{fig:opt}
\end{figure*}

\noindent 
Here, $\hat{\nu}$ denotes the unit vector along the $\nu$-direction (i.e., $\hat{x} = (1, 0)$, $\hat{y} = (0, 1)$) and the quasiparticle wavefunction of the $m^{\rm{th}}$ eigen state for site $\mathbf{i}$ takes the form $\phi^{m}_{\mathbf{i}} = \left[u^{m}_{\mathbf{i}\uparrow},\, u^{m}_{\mathbf{i}\downarrow},\, v^{m}_{\mathbf{i}\downarrow},\, v^{m}_{\mathbf{n}\uparrow}\right]^{\rm{T}}$. Note that, $t^{\prime}$ represents all generalized hoppings for a given bond. After computing the bond currents for all bonds, we average them to obtain the system’s net supercurrent, $\mathbf{I}$. This supercurrent can be expressed in polar form as $\mathbf{I}=(I\cos(\alpha), I\sin(\alpha))$, where $I=|\mathbf{I}|$ is the magnitude and $\alpha$ denotes the corresponding polar angle. We demonstrate SDE by examining the behavior of the supercurrent density $\mathbf{I}(\mathbf{q})$ for various system parameters considering the lattice Hamiltonian $H_l$ [Eq.~\eqref{eq:lattice1}] as well as the effective regularized Hamiltonian $H^{\prime}_l$ [Eq.~(\ref{eq:ham_real})]. The magnitude of the supercurrent density in the $q_x$–$q_y$ plane is illustrated in Fig.~\ref{fig:current}. In the absence of the conical spin texture ($J=0$), $|\mathbf{I}|$ remains symmetric with respect to the Cooper pair momentum $\mathbf{q}$, as shown in Fig.~\ref{fig:current}(a),(c). The supercurrent density vanishes beyond a critical momentum $q_c$, consistent with the suppression of the superconducting order parameter $\Delta$ beyond this threshold. Remarkably, when a spin texture corresponding to the cone phase is introduced via a finite exchange interaction ($J \ne 0$), $|\mathbf{I}|$ develops a clear asymmetry in $\mathbf{q}$ space, as evident from Fig.~\ref{fig:current}(b),(d). This asymmetry arises due to the simultaneous breaking of inversion and time-reversal symmetries by the spin texture, along with the emergence of the FFLO superconducting phase. The latter is characterized by a finite ground-state Cooper pair momentum $\mathbf{q}_0$ where the supercurrent vanishes, i.e., $\mathbf{I}(\mathbf{q}_0) = 0$. This asymmetry is particularly pronounced along the diagonal direction $q_x = q_y$ (see Fig.~\ref{fig:current}(b),(d)), owing to the symmetric choice of the 
spin texture pitch vector, $g_x = g_y$.


Then, we identify the critical current $I_c$ by maximizing $I$ for each value of $\alpha$. The corresponding efficiency is then defined as~\cite{Sayan_2024}:
\begin{equation}
    \eta(\alpha) = \frac{I_c(\alpha)-I_c(\alpha + \pi)}{I_c(\alpha)+I_c(\alpha + \pi)}\ .
\end{equation}

The angular dependence of the critical current $I_c$ is shown in Fig.~\ref{fig:eff}(a,b). In the absence of the spin texture ($J = 0$), the critical current remains symmetric, satisfying $I_c(\alpha) = I_c(\alpha + \pi)$, consistent with the absence of FFLO pairing when $J=0$. However, when a finite exchange coupling is introduced ($J \ne 0$), the critical current exhibits a clear non-reciprocal behavior resulting  in SDE with finite diode efficiency. As shown in Fig.~\ref{fig:eff}(c,d), the diode efficiency $\eta$ displays a distinct angular dependence. Notably, $\eta$ reaches a maximum around $\alpha = \pi/4$ (diagonal direction), then gradually decreases and crosses zero at $\alpha=3\pi/4$ with a reversal in sign of efficiency. This pronounced angular dependence can be directly attributed to the underlying symmetries of the magnetic texture configuration. For the symmetric choice of the pitch vector $g_x = g_y$, the spin texture propagates along the diagonal direction ($\alpha = \pi/4$), where the critical current $I_c$ exhibits maximum asymmetry. On the contrary, for $\alpha = 3\pi/4$, the system preserves inversion symmetry perpendicular to the propagation direction of the spin texture, which is reflected in the Bogoliubov–de Gennes (BdG) band structure that satisfies $H(\mathbf{k}) = H(-\mathbf{k})$ along the $k_x = -k_y$ line. This symmetry prevents the emergence of non-reciprocal behavior essential for the  SDE (see the supplementary material (SM) for detailed discussion). For $J=0$, the efficiency $\eta$ vanishes for any value of $\alpha$. We also explore the angular dependence of $I_c$ and $\eta$ for planar helical and antiferromagnetic spin textures, where an external magnetic field is required to break the symmetries necessary for realizing the SDE. The planar helical texture achieves a peak efficiency of around $20\%$, while the antiferromagnetic case yields only about $3\%$ (see SM for details).

Afterwards, we explore the optimization of the diode efficiency $\eta$ by systematically varying the key system parameters of the microscopic
lattice Hamiltonian described by Eq.~(\ref{eq:lattice1}). This analysis identifies the parameter regimes that yield maximum diode efficiency and reveals how $\eta$ depends sensitively on the underlying characteristics of our system [see Fig.~\ref{fig:opt}]. In Fig.~\ref{fig:opt}(a), $\eta$ is illustrated as a function of the cone angle of the spin texture for different values of chemical potential $\mu$, highlighting finite diode efficiency within the range $0<\theta<\pi/2$. As shown in Fig.~\ref{fig:opt}(a), the diode efficiency $\eta$ vanishes for the trivial or planar spin texture configurations ($\theta = 0, \pi/2$).  These configurations unable to support a finite FFLO ground state Cooper pair momentum $q_0$, as discussed in earlier sections [see Fig.~\ref{fig:self}(d,h)]. When investigated as a function of chemical potential $\mu$ for different values of $\theta$, the diode efficiency $\eta$ exhibits a characteristic behavior: it increases gradually from zero at $\mu = 0$, reaches a sharp peak at finite $\mu$, and then rapidly drops, as illustrated in Fig.~\ref{fig:opt}(b). Notably, $\eta$ also changes sign as the polarity of $\mu$ is reversed. The optimal value of $\mu$ at which the efficiency reaches a peaked maximum, also depends sensitively on other system parameters, such as the exchange coupling $J$ and the cone angle $\theta$. For all the results presented in Fig.~\ref{fig:opt}, we fix $\alpha = \pi/4$, as the diode efficiency becomes most pronounced at this angle for the chosen spin texture configuration, as discussed in earlier sections [see Fig.~\ref{fig:eff}]. 
 
As evident from Fig.~\ref{fig:opt}(c), the diode efficiency $\eta$ begins to develop significantly for $J/\Delta_0 > 0.3$, with a pronounced enhancement near $J/\Delta_0 = 0.5$, where it reaches an optimal efficiency of approximately 50$\%$. It is important to emphasize that a higher diode efficiency does not necessarily imply larger asymmetry in the critical current: $\Delta I_c = I_c(\alpha) - I_c(\alpha + \pi)$. This distinction is crucial, both for practical applications concerning SDE and for understanding recent experimental observations where, a higher asymmetry in critical currents is often desired~\cite{lin_2022}. Although, $\eta$ continues to grow remarkably beyond $J/\Delta_0 = 0.5$, the critical current asymmetry $\Delta I_c$ peaks around this value and sharply decreases for larger $J/\Delta_0$, as shown in the inset of Fig.~\ref{fig:opt}(c). A similar trend is observed when $\Delta I_c$ is investigated as a function of chemical potential $\mu$, where the asymmetry peaks near $\mu/\Delta_0 = 1$ [see inset of Fig.~\ref{fig:opt}(b)]. Therefore, based on these observations, we identify the set of parameters $(J/\Delta_0, \mu/\Delta_0, \alpha) = (0.5, 1, \pi/4)$ to demonstrate optimization of the diode performance of our device.  A similar optimization analysis for the effective lattice-regularized model is presented in detail in the SM.

\section{Summary and discussion} \label{sec6}

To summarize, we present a potential realization of the SDE in a simple yet versatile platform: a classical conical magnetic texture deposited on a conventional bulk $s$-wave superconductor, requiring no external fields. Using a self-consistent BdG mean-field analysis of both the microscopic lattice Hamiltonian [Eq.(\ref{eq:lattice1})] and its effective lattice-regularized low-energy counterpart [Eq.(\ref{eq:ham_real})], we establish the emergence of an FFLO superconducting ground state with a unique Cooper pair momentum $q_0$. Results from the corresponding continuum (momentum-space) model [Eq.~(\ref{eq:k_space_hamiltonian})] are provided in the SM. The conical spin texture ($0 < \theta < \pi/2$) enables this field-free SDE by simultaneously breaking time-reversal symmetry through in-plane spin winding and inversion symmetry via its finite out-of-plane component, leading to non-reciprocal critical currents. We also optimize the diode efficiency across different spin-texture configurations. The cone-phase texture achieves efficiencies up to $50\%$, while a planar helical texture ($\theta = \pi/2$) can produce comparable performance only under an external Zeeman field perpendicular to the spiral plane. In contrast, the efficiency drops sharply for an antiferromagnetic texture, which, like the planar case, requires an external Zeeman field to exhibit the SDE (see SM for details).

In recent times, the potential realizations of non-collinear spin textures on superconducting substrates have gained significant attention, particularly in setups concerning magnet–superconductor hybrid systems. Very recently, spin-polarized scanning tunneling microscopy (SP-STM) studies have revealed a stabilized spin-spiral ground state with a magnetic periodicity of approximately~6\,nm in a monolayer of Fe deposited on a Ta(110) surface~\cite{Weis,Roland}. The underlying Ta(110), a conventional $s$-wave superconductor with a critical temperature of approximately~$4.5$\,K and a sizable superconducting gap in the range of $0.7$--$0.9$\,meV, closely aligns with our model assumptions, where we consider $\Delta_0 = 1$\,meV. This non-collinear state is stabilized through a combination of competing exchange interactions owing to effects of frustration. First-principles calculations using the relativistic Korringa–Kohn–Rostoker method~\cite{DFT} reveal that such spiral texture is stabilized by the interplay of strong ferromagnetic nearest-neighbor exchange  and competing antiferromagnetic further-neighbor couplings ($|J| <\approx 27$\,meV), all mediated by the strong spin–orbit coupling (SOC) intrinsic to the Ta substrate. These findings demonstrate the viability of stabilizing robust non-collinear magnetic textures in atomically thin magnetic films interfaced with conventional superconductors, thereby highlighting the experimental accessibility and practical relevance of our proposed model for realizing a highly efficient superconducting diode in 2D magnet–superconductor hybrid structure.

\section{Acknowledgments} 
SKG acknowledges financial support from Anusandhan National Research Foundation (ANRF) erstwhile Science and Engineering Research Board (SERB), Government of India via the Startup Research Grant: SRG/2023/000934 and from IIT Kanpur via the Initiation Grant (IITK/PHY/2022116). DS and SKG utilized the \textit{Andromeda} server at IIT Kanpur for numerical calculations. SB and AS acknowledge the SAMKHYA: HPC Facility provided at IOP, Bhubaneswar and the two workstations provided by IOP, Bhubaneshwar from the DAE APEX project for the numerical computations.

\bibliography{refs}

\begin{thebibliography}{60}%
\makeatletter
\providecommand \@ifxundefined [1]{%
 \@ifx{#1\undefined}
}%
\providecommand \@ifnum [1]{%
 \ifnum #1\expandafter \@firstoftwo
 \else \expandafter \@secondoftwo
 \fi
}%
\providecommand \@ifx [1]{%
 \ifx #1\expandafter \@firstoftwo
 \else \expandafter \@secondoftwo
 \fi
}%
\providecommand \natexlab [1]{#1}%
\providecommand \enquote  [1]{``#1''}%
\providecommand \bibnamefont  [1]{#1}%
\providecommand \bibfnamefont [1]{#1}%
\providecommand \citenamefont [1]{#1}%
\providecommand \href@noop [0]{\@secondoftwo}%
\providecommand \href [0]{\begingroup \@sanitize@url \@href}%
\providecommand \@href[1]{\@@startlink{#1}\@@href}%
\providecommand \@@href[1]{\endgroup#1\@@endlink}%
\providecommand \@sanitize@url [0]{\catcode `\\12\catcode `\$12\catcode
  `\&12\catcode `\#12\catcode `\^12\catcode `\_12\catcode `\%12\relax}%
\providecommand \@@startlink[1]{}%
\providecommand \@@endlink[0]{}%
\providecommand \url  [0]{\begingroup\@sanitize@url \@url }%
\providecommand \@url [1]{\endgroup\@href {#1}{\urlprefix }}%
\providecommand \urlprefix  [0]{URL }%
\providecommand \Eprint [0]{\href }%
\providecommand \doibase [0]{https://doi.org/}%
\providecommand \selectlanguage [0]{\@gobble}%
\providecommand \bibinfo  [0]{\@secondoftwo}%
\providecommand \bibfield  [0]{\@secondoftwo}%
\providecommand \translation [1]{[#1]}%
\providecommand \BibitemOpen [0]{}%
\providecommand \bibitemStop [0]{}%
\providecommand \bibitemNoStop [0]{.\EOS\space}%
\providecommand \EOS [0]{\spacefactor3000\relax}%
\providecommand \BibitemShut  [1]{\csname bibitem#1\endcsname}%
\let\auto@bib@innerbib\@empty
\bibitem [{\citenamefont {Nadeem}\ \emph {et~al.}(2023)\citenamefont {Nadeem},
  \citenamefont {Fuhrer},\ and\ \citenamefont {Wang}}]{nadeem_2023}%
  \BibitemOpen
  \bibfield  {author} {\bibinfo {author} {\bibfnamefont {M.}~\bibnamefont
  {Nadeem}}, \bibinfo {author} {\bibfnamefont {M.~S.}\ \bibnamefont {Fuhrer}},\
  and\ \bibinfo {author} {\bibfnamefont {X.}~\bibnamefont {Wang}},\ }\bibfield
  {title} {\bibinfo {title} {The superconducting diode effect},\ }\href
  {https://doi.org/10.1038/s42254-023-00632-w} {\bibfield  {journal} {\bibinfo
  {journal} {Nature Reviews Physics}\ }\textbf {\bibinfo {volume} {5}},\
  \bibinfo {pages} {558} (\bibinfo {year} {2023})}\BibitemShut {NoStop}%
\bibitem [{\citenamefont {Nagaosa}\ and\ \citenamefont
  {Yanase}(2024)}]{Nagaosa_2024}%
  \BibitemOpen
  \bibfield  {author} {\bibinfo {author} {\bibfnamefont {N.}~\bibnamefont
  {Nagaosa}}\ and\ \bibinfo {author} {\bibfnamefont {Y.}~\bibnamefont
  {Yanase}},\ }\bibfield  {title} {\bibinfo {title} {Nonreciprocal transport
  and optical phenomena in quantum materials},\ }\href
  {https://doi.org/https://doi.org/10.1146/annurev-conmatphys-032822-033734}
  {\bibfield  {journal} {\bibinfo  {journal} {Annual Review of Condensed Matter
  Physics}\ }\textbf {\bibinfo {volume} {15}},\ \bibinfo {pages} {63} (\bibinfo
  {year} {2024})}\BibitemShut {NoStop}%
\bibitem [{\citenamefont {Broussard}\ and\ \citenamefont
  {Geballe}(1988)}]{Broussard_1988}%
  \BibitemOpen
  \bibfield  {author} {\bibinfo {author} {\bibfnamefont {P.~R.}\ \bibnamefont
  {Broussard}}\ and\ \bibinfo {author} {\bibfnamefont {T.~H.}\ \bibnamefont
  {Geballe}},\ }\bibfield  {title} {\bibinfo {title} {Critical currents in
  sputtered nb-ta multilayers},\ }\href
  {https://doi.org/10.1103/PhysRevB.37.68} {\bibfield  {journal} {\bibinfo
  {journal} {Phys. Rev. B}\ }\textbf {\bibinfo {volume} {37}},\ \bibinfo
  {pages} {68} (\bibinfo {year} {1988})}\BibitemShut {NoStop}%
\bibitem [{\citenamefont {Jiang}\ \emph {et~al.}(1994)\citenamefont {Jiang},
  \citenamefont {Connolly}, \citenamefont {Hagen},\ and\ \citenamefont
  {Lobb}}]{Jiang_1994}%
  \BibitemOpen
  \bibfield  {author} {\bibinfo {author} {\bibfnamefont {X.}~\bibnamefont
  {Jiang}}, \bibinfo {author} {\bibfnamefont {P.~J.}\ \bibnamefont {Connolly}},
  \bibinfo {author} {\bibfnamefont {S.~J.}\ \bibnamefont {Hagen}},\ and\
  \bibinfo {author} {\bibfnamefont {C.~J.}\ \bibnamefont {Lobb}},\ }\bibfield
  {title} {\bibinfo {title} {Asymmetric current-voltage characteristics in
  type-ii superconductors},\ }\href {https://doi.org/10.1103/PhysRevB.49.9244}
  {\bibfield  {journal} {\bibinfo  {journal} {Phys. Rev. B}\ }\textbf {\bibinfo
  {volume} {49}},\ \bibinfo {pages} {9244} (\bibinfo {year}
  {1994})}\BibitemShut {NoStop}%
\bibitem [{\citenamefont {Papon}\ \emph {et~al.}(2008)\citenamefont {Papon},
  \citenamefont {Senapati},\ and\ \citenamefont {Barber}}]{Papon_2008}%
  \BibitemOpen
  \bibfield  {author} {\bibinfo {author} {\bibfnamefont {A.}~\bibnamefont
  {Papon}}, \bibinfo {author} {\bibfnamefont {K.}~\bibnamefont {Senapati}},\
  and\ \bibinfo {author} {\bibfnamefont {Z.~H.}\ \bibnamefont {Barber}},\
  }\bibfield  {title} {\bibinfo {title} {Asymmetric critical current of niobium
  microbridges with ferromagnetic stripe},\ }\href
  {https://doi.org/10.1063/1.3009207} {\bibfield  {journal} {\bibinfo
  {journal} {Applied Physics Letters}\ }\textbf {\bibinfo {volume} {93}},\
  \bibinfo {pages} {172507} (\bibinfo {year} {2008})}\BibitemShut {NoStop}%
\bibitem [{\citenamefont {Narita}\ and\ \citenamefont
  {Ono}(2024)}]{Narita_2024}%
  \BibitemOpen
  \bibfield  {author} {\bibinfo {author} {\bibfnamefont {H.}~\bibnamefont
  {Narita}}\ and\ \bibinfo {author} {\bibfnamefont {T.}~\bibnamefont {Ono}},\
  }\bibfield  {title} {\bibinfo {title} {Superconducting diode effect in
  artificial superlattices},\ }\href {https://doi.org/10.11470/jsaprev.240206}
  {\bibfield  {journal} {\bibinfo  {journal} {JSAP Review}\ }\textbf {\bibinfo
  {volume} {2024}},\ \bibinfo {pages} {240206} (\bibinfo {year}
  {2024})}\BibitemShut {NoStop}%
\bibitem [{\citenamefont {Ando}\ \emph {et~al.}(2020)\citenamefont {Ando},
  \citenamefont {Miyasaka}, \citenamefont {Li}, \citenamefont {Ishizuka},
  \citenamefont {Arakawa}, \citenamefont {Shiota}, \citenamefont {Moriyama},
  \citenamefont {Yanase},\ and\ \citenamefont {Ono}}]{ando_2020}%
  \BibitemOpen
  \bibfield  {author} {\bibinfo {author} {\bibfnamefont {F.}~\bibnamefont
  {Ando}}, \bibinfo {author} {\bibfnamefont {Y.}~\bibnamefont {Miyasaka}},
  \bibinfo {author} {\bibfnamefont {T.}~\bibnamefont {Li}}, \bibinfo {author}
  {\bibfnamefont {J.}~\bibnamefont {Ishizuka}}, \bibinfo {author}
  {\bibfnamefont {T.}~\bibnamefont {Arakawa}}, \bibinfo {author} {\bibfnamefont
  {Y.}~\bibnamefont {Shiota}}, \bibinfo {author} {\bibfnamefont
  {T.}~\bibnamefont {Moriyama}}, \bibinfo {author} {\bibfnamefont
  {Y.}~\bibnamefont {Yanase}},\ and\ \bibinfo {author} {\bibfnamefont
  {T.}~\bibnamefont {Ono}},\ }\bibfield  {title} {\bibinfo {title} {Observation
  of superconducting diode effect},\ }\href
  {https://doi.org/10.1038/s41586-020-2590-4} {\bibfield  {journal} {\bibinfo
  {journal} {Nature}\ }\textbf {\bibinfo {volume} {584}},\ \bibinfo {pages}
  {373} (\bibinfo {year} {2020})}\BibitemShut {NoStop}%
\bibitem [{\citenamefont {Sundaresh}\ \emph {et~al.}(2023)\citenamefont
  {Sundaresh}, \citenamefont {V{\"a}yrynen}, \citenamefont {Lyanda-Geller},\
  and\ \citenamefont {Rokhinson}}]{sundaresh_2023}%
  \BibitemOpen
  \bibfield  {author} {\bibinfo {author} {\bibfnamefont {A.}~\bibnamefont
  {Sundaresh}}, \bibinfo {author} {\bibfnamefont {J.~I.}\ \bibnamefont
  {V{\"a}yrynen}}, \bibinfo {author} {\bibfnamefont {Y.}~\bibnamefont
  {Lyanda-Geller}},\ and\ \bibinfo {author} {\bibfnamefont {L.~P.}\
  \bibnamefont {Rokhinson}},\ }\bibfield  {title} {\bibinfo {title}
  {Diamagnetic mechanism of critical current non-reciprocity in multilayered
  superconductors},\ }\href {https://doi.org/10.1038/s41467-023-36786-5}
  {\bibfield  {journal} {\bibinfo  {journal} {Nature Communications}\ }\textbf
  {\bibinfo {volume} {14}},\ \bibinfo {pages} {1628} (\bibinfo {year}
  {2023})}\BibitemShut {NoStop}%
\bibitem [{\citenamefont {Wakatsuki}\ \emph {et~al.}(2017)\citenamefont
  {Wakatsuki}, \citenamefont {Saito}, \citenamefont {Hoshino}, \citenamefont
  {Itahashi}, \citenamefont {Ideue}, \citenamefont {Ezawa}, \citenamefont
  {Iwasa},\ and\ \citenamefont {Nagaosa}}]{Wakatsuki_2017}%
  \BibitemOpen
  \bibfield  {author} {\bibinfo {author} {\bibfnamefont {R.}~\bibnamefont
  {Wakatsuki}}, \bibinfo {author} {\bibfnamefont {Y.}~\bibnamefont {Saito}},
  \bibinfo {author} {\bibfnamefont {S.}~\bibnamefont {Hoshino}}, \bibinfo
  {author} {\bibfnamefont {Y.~M.}\ \bibnamefont {Itahashi}}, \bibinfo {author}
  {\bibfnamefont {T.}~\bibnamefont {Ideue}}, \bibinfo {author} {\bibfnamefont
  {M.}~\bibnamefont {Ezawa}}, \bibinfo {author} {\bibfnamefont
  {Y.}~\bibnamefont {Iwasa}},\ and\ \bibinfo {author} {\bibfnamefont
  {N.}~\bibnamefont {Nagaosa}},\ }\bibfield  {title} {\bibinfo {title}
  {Nonreciprocal charge transport in noncentrosymmetric superconductors},\
  }\href {https://doi.org/10.1126/sciadv.1602390} {\bibfield  {journal}
  {\bibinfo  {journal} {Science Advances}\ }\textbf {\bibinfo {volume} {3}},\
  \bibinfo {pages} {e1602390} (\bibinfo {year} {2017})}\BibitemShut {NoStop}%
\bibitem [{\citenamefont {Itahashi}\ \emph {et~al.}(2020)\citenamefont
  {Itahashi}, \citenamefont {Ideue}, \citenamefont {Saito}, \citenamefont
  {Shimizu}, \citenamefont {Ouchi}, \citenamefont {Nojima},\ and\ \citenamefont
  {Iwasa}}]{Yuki_2020}%
  \BibitemOpen
  \bibfield  {author} {\bibinfo {author} {\bibfnamefont {Y.~M.}\ \bibnamefont
  {Itahashi}}, \bibinfo {author} {\bibfnamefont {T.}~\bibnamefont {Ideue}},
  \bibinfo {author} {\bibfnamefont {Y.}~\bibnamefont {Saito}}, \bibinfo
  {author} {\bibfnamefont {S.}~\bibnamefont {Shimizu}}, \bibinfo {author}
  {\bibfnamefont {T.}~\bibnamefont {Ouchi}}, \bibinfo {author} {\bibfnamefont
  {T.}~\bibnamefont {Nojima}},\ and\ \bibinfo {author} {\bibfnamefont
  {Y.}~\bibnamefont {Iwasa}},\ }\bibfield  {title} {\bibinfo {title}
  {Nonreciprocal transport in gate-induced polar superconductor
  $\mathrm{SrTiO}_{3}$},\ }\href {https://doi.org/10.1126/sciadv.aay9120}
  {\bibfield  {journal} {\bibinfo  {journal} {Science Advances}\ }\textbf
  {\bibinfo {volume} {6}},\ \bibinfo {pages} {eaay9120} (\bibinfo {year}
  {2020})}\BibitemShut {NoStop}%
\bibitem [{\citenamefont {Schumann}\ \emph {et~al.}(2020)\citenamefont
  {Schumann}, \citenamefont {Galletti}, \citenamefont {Jeong}, \citenamefont
  {Ahadi}, \citenamefont {Strickland}, \citenamefont {Salmani-Rezaie},\ and\
  \citenamefont {Stemmer}}]{Schumann_2020}%
  \BibitemOpen
  \bibfield  {author} {\bibinfo {author} {\bibfnamefont {T.}~\bibnamefont
  {Schumann}}, \bibinfo {author} {\bibfnamefont {L.}~\bibnamefont {Galletti}},
  \bibinfo {author} {\bibfnamefont {H.}~\bibnamefont {Jeong}}, \bibinfo
  {author} {\bibfnamefont {K.}~\bibnamefont {Ahadi}}, \bibinfo {author}
  {\bibfnamefont {W.~M.}\ \bibnamefont {Strickland}}, \bibinfo {author}
  {\bibfnamefont {S.}~\bibnamefont {Salmani-Rezaie}},\ and\ \bibinfo {author}
  {\bibfnamefont {S.}~\bibnamefont {Stemmer}},\ }\bibfield  {title} {\bibinfo
  {title} {Possible signatures of mixed-parity superconductivity in doped polar
  $\mathrm{SrTi}{\mathrm{o}}_{3}$ films},\ }\href
  {https://doi.org/10.1103/PhysRevB.101.100503} {\bibfield  {journal} {\bibinfo
   {journal} {Phys. Rev. B}\ }\textbf {\bibinfo {volume} {101}},\ \bibinfo
  {pages} {100503} (\bibinfo {year} {2020})}\BibitemShut {NoStop}%
\bibitem [{\citenamefont {Lin}\ \emph {et~al.}(2022)\citenamefont {Lin},
  \citenamefont {Siriviboon}, \citenamefont {Scammell}, \citenamefont {Liu},
  \citenamefont {Rhodes}, \citenamefont {Watanabe}, \citenamefont {Taniguchi},
  \citenamefont {Hone}, \citenamefont {Scheurer},\ and\ \citenamefont
  {Li}}]{lin_2022}%
  \BibitemOpen
  \bibfield  {author} {\bibinfo {author} {\bibfnamefont {J.-X.}\ \bibnamefont
  {Lin}}, \bibinfo {author} {\bibfnamefont {P.}~\bibnamefont {Siriviboon}},
  \bibinfo {author} {\bibfnamefont {H.~D.}\ \bibnamefont {Scammell}}, \bibinfo
  {author} {\bibfnamefont {S.}~\bibnamefont {Liu}}, \bibinfo {author}
  {\bibfnamefont {D.}~\bibnamefont {Rhodes}}, \bibinfo {author} {\bibfnamefont
  {K.}~\bibnamefont {Watanabe}}, \bibinfo {author} {\bibfnamefont
  {T.}~\bibnamefont {Taniguchi}}, \bibinfo {author} {\bibfnamefont
  {J.}~\bibnamefont {Hone}}, \bibinfo {author} {\bibfnamefont {M.~S.}\
  \bibnamefont {Scheurer}},\ and\ \bibinfo {author} {\bibfnamefont
  {J.}~\bibnamefont {Li}},\ }\bibfield  {title} {\bibinfo {title} {Zero-field
  superconducting diode effect in small-twist-angle trilayer graphene},\ }\href
  {https://doi.org/10.1038/s41567-022-01700-1} {\bibfield  {journal} {\bibinfo
  {journal} {Nature Physics}\ }\textbf {\bibinfo {volume} {18}},\ \bibinfo
  {pages} {1221} (\bibinfo {year} {2022})}\BibitemShut {NoStop}%
\bibitem [{\citenamefont {Diez-Merida}\ \emph {et~al.}(2023)\citenamefont
  {Diez-Merida}, \citenamefont {D{\'\i}ez-Carl{\'o}n}, \citenamefont {Yang},
  \citenamefont {Xie}, \citenamefont {Gao}, \citenamefont {Senior},
  \citenamefont {Watanabe}, \citenamefont {Taniguchi}, \citenamefont {Lu},
  \citenamefont {Higginbotham} \emph {et~al.}}]{Jaime_2023}%
  \BibitemOpen
  \bibfield  {author} {\bibinfo {author} {\bibfnamefont {J.}~\bibnamefont
  {Diez-Merida}}, \bibinfo {author} {\bibfnamefont {A.}~\bibnamefont
  {D{\'\i}ez-Carl{\'o}n}}, \bibinfo {author} {\bibfnamefont {S.}~\bibnamefont
  {Yang}}, \bibinfo {author} {\bibfnamefont {Y.-M.}\ \bibnamefont {Xie}},
  \bibinfo {author} {\bibfnamefont {X.-J.}\ \bibnamefont {Gao}}, \bibinfo
  {author} {\bibfnamefont {J.}~\bibnamefont {Senior}}, \bibinfo {author}
  {\bibfnamefont {K.}~\bibnamefont {Watanabe}}, \bibinfo {author}
  {\bibfnamefont {T.}~\bibnamefont {Taniguchi}}, \bibinfo {author}
  {\bibfnamefont {X.}~\bibnamefont {Lu}}, \bibinfo {author} {\bibfnamefont
  {A.~P.}\ \bibnamefont {Higginbotham}}, \emph {et~al.},\ }\bibfield  {title}
  {\bibinfo {title} {Symmetry-broken josephson junctions and superconducting
  diodes in magic-angle twisted bilayer graphene},\ }\href
  {https://doi.org/10.1038/s41467-023-38005-7} {\bibfield  {journal} {\bibinfo
  {journal} {Nature Communications}\ }\textbf {\bibinfo {volume} {14}},\
  \bibinfo {pages} {2396} (\bibinfo {year} {2023})}\BibitemShut {NoStop}%
\bibitem [{\citenamefont {Bauriedl}\ \emph {et~al.}(2022)\citenamefont
  {Bauriedl}, \citenamefont {B{\"a}uml}, \citenamefont {Fuchs}, \citenamefont
  {Baumgartner}, \citenamefont {Paulik}, \citenamefont {Bauer}, \citenamefont
  {Lin}, \citenamefont {Lupton}, \citenamefont {Taniguchi}, \citenamefont
  {Watanabe} \emph {et~al.}}]{bauriedl_2022}%
  \BibitemOpen
  \bibfield  {author} {\bibinfo {author} {\bibfnamefont {L.}~\bibnamefont
  {Bauriedl}}, \bibinfo {author} {\bibfnamefont {C.}~\bibnamefont {B{\"a}uml}},
  \bibinfo {author} {\bibfnamefont {L.}~\bibnamefont {Fuchs}}, \bibinfo
  {author} {\bibfnamefont {C.}~\bibnamefont {Baumgartner}}, \bibinfo {author}
  {\bibfnamefont {N.}~\bibnamefont {Paulik}}, \bibinfo {author} {\bibfnamefont
  {J.~M.}\ \bibnamefont {Bauer}}, \bibinfo {author} {\bibfnamefont {K.-Q.}\
  \bibnamefont {Lin}}, \bibinfo {author} {\bibfnamefont {J.~M.}\ \bibnamefont
  {Lupton}}, \bibinfo {author} {\bibfnamefont {T.}~\bibnamefont {Taniguchi}},
  \bibinfo {author} {\bibfnamefont {K.}~\bibnamefont {Watanabe}}, \emph
  {et~al.},\ }\bibfield  {title} {\bibinfo {title} {Supercurrent diode effect
  and magnetochiral anisotropy in few-layer nbse2},\ }\href
  {https://doi.org/10.1038/s41467-022-31954-5} {\bibfield  {journal} {\bibinfo
  {journal} {Nature communications}\ }\textbf {\bibinfo {volume} {13}},\
  \bibinfo {pages} {4266} (\bibinfo {year} {2022})}\BibitemShut {NoStop}%
\bibitem [{\citenamefont {Yun}\ \emph {et~al.}(2023)\citenamefont {Yun},
  \citenamefont {Son}, \citenamefont {Shin}, \citenamefont {Park},
  \citenamefont {Zhang}, \citenamefont {Shin}, \citenamefont {Park},\ and\
  \citenamefont {Kim}}]{Yun_2023}%
  \BibitemOpen
  \bibfield  {author} {\bibinfo {author} {\bibfnamefont {J.}~\bibnamefont
  {Yun}}, \bibinfo {author} {\bibfnamefont {S.}~\bibnamefont {Son}}, \bibinfo
  {author} {\bibfnamefont {J.}~\bibnamefont {Shin}}, \bibinfo {author}
  {\bibfnamefont {G.}~\bibnamefont {Park}}, \bibinfo {author} {\bibfnamefont
  {K.}~\bibnamefont {Zhang}}, \bibinfo {author} {\bibfnamefont {Y.~J.}\
  \bibnamefont {Shin}}, \bibinfo {author} {\bibfnamefont {J.-G.}\ \bibnamefont
  {Park}},\ and\ \bibinfo {author} {\bibfnamefont {D.}~\bibnamefont {Kim}},\
  }\bibfield  {title} {\bibinfo {title} {Magnetic proximity-induced
  superconducting diode effect and infinite magnetoresistance in a van der
  waals heterostructure},\ }\href
  {https://doi.org/10.1103/PhysRevResearch.5.L022064} {\bibfield  {journal}
  {\bibinfo  {journal} {Phys. Rev. Res.}\ }\textbf {\bibinfo {volume} {5}},\
  \bibinfo {pages} {L022064} (\bibinfo {year} {2023})}\BibitemShut {NoStop}%
\bibitem [{\citenamefont {Zhang}\ \emph {et~al.}(2022)\citenamefont {Zhang},
  \citenamefont {Gu}, \citenamefont {Li}, \citenamefont {Hu},\ and\
  \citenamefont {Jiang}}]{Zhang_2022}%
  \BibitemOpen
  \bibfield  {author} {\bibinfo {author} {\bibfnamefont {Y.}~\bibnamefont
  {Zhang}}, \bibinfo {author} {\bibfnamefont {Y.}~\bibnamefont {Gu}}, \bibinfo
  {author} {\bibfnamefont {P.}~\bibnamefont {Li}}, \bibinfo {author}
  {\bibfnamefont {J.}~\bibnamefont {Hu}},\ and\ \bibinfo {author}
  {\bibfnamefont {K.}~\bibnamefont {Jiang}},\ }\bibfield  {title} {\bibinfo
  {title} {General theory of josephson diodes},\ }\href
  {https://doi.org/10.1103/PhysRevX.12.041013} {\bibfield  {journal} {\bibinfo
  {journal} {Phys. Rev. X}\ }\textbf {\bibinfo {volume} {12}},\ \bibinfo
  {pages} {041013} (\bibinfo {year} {2022})}\BibitemShut {NoStop}%
\bibitem [{\citenamefont {Kokkeler}\ \emph {et~al.}(2022)\citenamefont
  {Kokkeler}, \citenamefont {Golubov},\ and\ \citenamefont
  {Bergeret}}]{Kokkeler_2022}%
  \BibitemOpen
  \bibfield  {author} {\bibinfo {author} {\bibfnamefont {T.~H.}\ \bibnamefont
  {Kokkeler}}, \bibinfo {author} {\bibfnamefont {A.~A.}\ \bibnamefont
  {Golubov}},\ and\ \bibinfo {author} {\bibfnamefont {F.~S.}\ \bibnamefont
  {Bergeret}},\ }\bibfield  {title} {\bibinfo {title} {Field-free anomalous
  junction and superconducting diode effect in spin-split
  superconductor/topological insulator junctions},\ }\href
  {https://doi.org/10.1103/PhysRevB.106.214504} {\bibfield  {journal} {\bibinfo
   {journal} {Phys. Rev. B}\ }\textbf {\bibinfo {volume} {106}},\ \bibinfo
  {pages} {214504} (\bibinfo {year} {2022})}\BibitemShut {NoStop}%
\bibitem [{\citenamefont {Tanaka}\ \emph {et~al.}(2022)\citenamefont {Tanaka},
  \citenamefont {Lu},\ and\ \citenamefont {Nagaosa}}]{Tanaka_2022}%
  \BibitemOpen
  \bibfield  {author} {\bibinfo {author} {\bibfnamefont {Y.}~\bibnamefont
  {Tanaka}}, \bibinfo {author} {\bibfnamefont {B.}~\bibnamefont {Lu}},\ and\
  \bibinfo {author} {\bibfnamefont {N.}~\bibnamefont {Nagaosa}},\ }\bibfield
  {title} {\bibinfo {title} {Theory of giant diode effect in $d$-wave
  superconductor junctions on the surface of a topological insulator},\ }\href
  {https://doi.org/10.1103/PhysRevB.106.214524} {\bibfield  {journal} {\bibinfo
   {journal} {Phys. Rev. B}\ }\textbf {\bibinfo {volume} {106}},\ \bibinfo
  {pages} {214524} (\bibinfo {year} {2022})}\BibitemShut {NoStop}%
\bibitem [{\citenamefont {Legg}\ \emph {et~al.}(2023)\citenamefont {Legg},
  \citenamefont {Laubscher}, \citenamefont {Loss},\ and\ \citenamefont
  {Klinovaja}}]{Legg_2023}%
  \BibitemOpen
  \bibfield  {author} {\bibinfo {author} {\bibfnamefont {H.~F.}\ \bibnamefont
  {Legg}}, \bibinfo {author} {\bibfnamefont {K.}~\bibnamefont {Laubscher}},
  \bibinfo {author} {\bibfnamefont {D.}~\bibnamefont {Loss}},\ and\ \bibinfo
  {author} {\bibfnamefont {J.}~\bibnamefont {Klinovaja}},\ }\bibfield  {title}
  {\bibinfo {title} {Parity-protected superconducting diode effect in
  topological josephson junctions},\ }\href
  {https://doi.org/10.1103/PhysRevB.108.214520} {\bibfield  {journal} {\bibinfo
   {journal} {Phys. Rev. B}\ }\textbf {\bibinfo {volume} {108}},\ \bibinfo
  {pages} {214520} (\bibinfo {year} {2023})}\BibitemShut {NoStop}%
\bibitem [{\citenamefont {Cuozzo}\ \emph {et~al.}(2024)\citenamefont {Cuozzo},
  \citenamefont {Pan}, \citenamefont {Shabani},\ and\ \citenamefont
  {Rossi}}]{Cuozzo_2024}%
  \BibitemOpen
  \bibfield  {author} {\bibinfo {author} {\bibfnamefont {J.~J.}\ \bibnamefont
  {Cuozzo}}, \bibinfo {author} {\bibfnamefont {W.}~\bibnamefont {Pan}},
  \bibinfo {author} {\bibfnamefont {J.}~\bibnamefont {Shabani}},\ and\ \bibinfo
  {author} {\bibfnamefont {E.}~\bibnamefont {Rossi}},\ }\bibfield  {title}
  {\bibinfo {title} {Microwave-tunable diode effect in asymmetric squids with
  topological josephson junctions},\ }\href
  {https://doi.org/10.1103/PhysRevResearch.6.023011} {\bibfield  {journal}
  {\bibinfo  {journal} {Phys. Rev. Res.}\ }\textbf {\bibinfo {volume} {6}},\
  \bibinfo {pages} {023011} (\bibinfo {year} {2024})}\BibitemShut {NoStop}%
\bibitem [{\citenamefont {Souto}\ \emph {et~al.}(2022)\citenamefont {Souto},
  \citenamefont {Leijnse},\ and\ \citenamefont {Schrade}}]{Souto_2022}%
  \BibitemOpen
  \bibfield  {author} {\bibinfo {author} {\bibfnamefont {R.~S.}\ \bibnamefont
  {Souto}}, \bibinfo {author} {\bibfnamefont {M.}~\bibnamefont {Leijnse}},\
  and\ \bibinfo {author} {\bibfnamefont {C.}~\bibnamefont {Schrade}},\
  }\bibfield  {title} {\bibinfo {title} {Josephson diode effect in supercurrent
  interferometers},\ }\href {https://doi.org/10.1103/PhysRevLett.129.267702}
  {\bibfield  {journal} {\bibinfo  {journal} {Phys. Rev. Lett.}\ }\textbf
  {\bibinfo {volume} {129}},\ \bibinfo {pages} {267702} (\bibinfo {year}
  {2022})}\BibitemShut {NoStop}%
\bibitem [{\citenamefont {Cheng}\ and\ \citenamefont {Sun}(2023)}]{Cheng_2023}%
  \BibitemOpen
  \bibfield  {author} {\bibinfo {author} {\bibfnamefont {Q.}~\bibnamefont
  {Cheng}}\ and\ \bibinfo {author} {\bibfnamefont {Q.-F.}\ \bibnamefont
  {Sun}},\ }\bibfield  {title} {\bibinfo {title} {Josephson diode based on
  conventional superconductors and a chiral quantum dot},\ }\href
  {https://doi.org/10.1103/PhysRevB.107.184511} {\bibfield  {journal} {\bibinfo
   {journal} {Phys. Rev. B}\ }\textbf {\bibinfo {volume} {107}},\ \bibinfo
  {pages} {184511} (\bibinfo {year} {2023})}\BibitemShut {NoStop}%
\bibitem [{\citenamefont {Steiner}\ \emph {et~al.}(2023)\citenamefont
  {Steiner}, \citenamefont {Melischek}, \citenamefont {Trahms}, \citenamefont
  {Franke},\ and\ \citenamefont {von Oppen}}]{Steiner_2023}%
  \BibitemOpen
  \bibfield  {author} {\bibinfo {author} {\bibfnamefont {J.~F.}\ \bibnamefont
  {Steiner}}, \bibinfo {author} {\bibfnamefont {L.}~\bibnamefont {Melischek}},
  \bibinfo {author} {\bibfnamefont {M.}~\bibnamefont {Trahms}}, \bibinfo
  {author} {\bibfnamefont {K.~J.}\ \bibnamefont {Franke}},\ and\ \bibinfo
  {author} {\bibfnamefont {F.}~\bibnamefont {von Oppen}},\ }\bibfield  {title}
  {\bibinfo {title} {Diode effects in current-biased josephson junctions},\
  }\href {https://doi.org/10.1103/PhysRevLett.130.177002} {\bibfield  {journal}
  {\bibinfo  {journal} {Phys. Rev. Lett.}\ }\textbf {\bibinfo {volume} {130}},\
  \bibinfo {pages} {177002} (\bibinfo {year} {2023})}\BibitemShut {NoStop}%
\bibitem [{\citenamefont {Costa}\ \emph {et~al.}(2023)\citenamefont {Costa},
  \citenamefont {Fabian},\ and\ \citenamefont {Kochan}}]{Costa_2023}%
  \BibitemOpen
  \bibfield  {author} {\bibinfo {author} {\bibfnamefont {A.}~\bibnamefont
  {Costa}}, \bibinfo {author} {\bibfnamefont {J.}~\bibnamefont {Fabian}},\ and\
  \bibinfo {author} {\bibfnamefont {D.}~\bibnamefont {Kochan}},\ }\bibfield
  {title} {\bibinfo {title} {Microscopic study of the josephson supercurrent
  diode effect in josephson junctions based on two-dimensional electron gas},\
  }\href {https://doi.org/10.1103/PhysRevB.108.054522} {\bibfield  {journal}
  {\bibinfo  {journal} {Phys. Rev. B}\ }\textbf {\bibinfo {volume} {108}},\
  \bibinfo {pages} {054522} (\bibinfo {year} {2023})}\BibitemShut {NoStop}%
\bibitem [{\citenamefont {Wei}\ \emph {et~al.}(2022)\citenamefont {Wei},
  \citenamefont {Liu}, \citenamefont {Wang},\ and\ \citenamefont
  {Liu}}]{Wei_2022}%
  \BibitemOpen
  \bibfield  {author} {\bibinfo {author} {\bibfnamefont {Y.-J.}\ \bibnamefont
  {Wei}}, \bibinfo {author} {\bibfnamefont {H.-L.}\ \bibnamefont {Liu}},
  \bibinfo {author} {\bibfnamefont {J.}~\bibnamefont {Wang}},\ and\ \bibinfo
  {author} {\bibfnamefont {J.-F.}\ \bibnamefont {Liu}},\ }\bibfield  {title}
  {\bibinfo {title} {Supercurrent rectification effect in graphene-based
  josephson junctions},\ }\href {https://doi.org/10.1103/PhysRevB.106.165419}
  {\bibfield  {journal} {\bibinfo  {journal} {Phys. Rev. B}\ }\textbf {\bibinfo
  {volume} {106}},\ \bibinfo {pages} {165419} (\bibinfo {year}
  {2022})}\BibitemShut {NoStop}%
\bibitem [{\citenamefont {Daido}\ and\ \citenamefont
  {Yanase}(2022)}]{Daido_2022}%
  \BibitemOpen
  \bibfield  {author} {\bibinfo {author} {\bibfnamefont {A.}~\bibnamefont
  {Daido}}\ and\ \bibinfo {author} {\bibfnamefont {Y.}~\bibnamefont {Yanase}},\
  }\bibfield  {title} {\bibinfo {title} {Superconducting diode effect and
  nonreciprocal transition lines},\ }\href
  {https://doi.org/10.1103/PhysRevB.106.205206} {\bibfield  {journal} {\bibinfo
   {journal} {Phys. Rev. B}\ }\textbf {\bibinfo {volume} {106}},\ \bibinfo
  {pages} {205206} (\bibinfo {year} {2022})}\BibitemShut {NoStop}%
\bibitem [{\citenamefont {Daido}\ \emph {et~al.}(2022)\citenamefont {Daido},
  \citenamefont {Ikeda},\ and\ \citenamefont {Yanase}}]{Daido_2022_intrinsic}%
  \BibitemOpen
  \bibfield  {author} {\bibinfo {author} {\bibfnamefont {A.}~\bibnamefont
  {Daido}}, \bibinfo {author} {\bibfnamefont {Y.}~\bibnamefont {Ikeda}},\ and\
  \bibinfo {author} {\bibfnamefont {Y.}~\bibnamefont {Yanase}},\ }\bibfield
  {title} {\bibinfo {title} {Intrinsic superconducting diode effect},\ }\href
  {https://doi.org/10.1103/PhysRevLett.128.037001} {\bibfield  {journal}
  {\bibinfo  {journal} {Phys. Rev. Lett.}\ }\textbf {\bibinfo {volume} {128}},\
  \bibinfo {pages} {037001} (\bibinfo {year} {2022})}\BibitemShut {NoStop}%
\bibitem [{\citenamefont {Yuan}\ and\ \citenamefont {Fu}(2022)}]{Yuan_2022}%
  \BibitemOpen
  \bibfield  {author} {\bibinfo {author} {\bibfnamefont {N.~F.~Q.}\
  \bibnamefont {Yuan}}\ and\ \bibinfo {author} {\bibfnamefont {L.}~\bibnamefont
  {Fu}},\ }\bibfield  {title} {\bibinfo {title} {Supercurrent diode effect and
  finite-momentum superconductors},\ }\href
  {https://doi.org/10.1073/pnas.2119548119} {\bibfield  {journal} {\bibinfo
  {journal} {Proceedings of the National Academy of Sciences}\ }\textbf
  {\bibinfo {volume} {119}},\ \bibinfo {pages} {e2119548119} (\bibinfo {year}
  {2022})}\BibitemShut {NoStop}%
\bibitem [{\citenamefont {He}\ \emph {et~al.}(2022)\citenamefont {He},
  \citenamefont {Tanaka},\ and\ \citenamefont {Nagaosa}}]{He_2022}%
  \BibitemOpen
  \bibfield  {author} {\bibinfo {author} {\bibfnamefont {J.~J.}\ \bibnamefont
  {He}}, \bibinfo {author} {\bibfnamefont {Y.}~\bibnamefont {Tanaka}},\ and\
  \bibinfo {author} {\bibfnamefont {N.}~\bibnamefont {Nagaosa}},\ }\bibfield
  {title} {\bibinfo {title} {A phenomenological theory of superconductor
  diodes},\ }\href {https://doi.org/10.1088/1367-2630/ac6766} {\bibfield
  {journal} {\bibinfo  {journal} {New Journal of Physics}\ }\textbf {\bibinfo
  {volume} {24}},\ \bibinfo {pages} {053014} (\bibinfo {year}
  {2022})}\BibitemShut {NoStop}%
\bibitem [{\citenamefont {Ili\ifmmode~\acute{c}\else \'{c}\fi{}}\ and\
  \citenamefont {Bergeret}(2022)}]{Ili_2022}%
  \BibitemOpen
  \bibfield  {author} {\bibinfo {author} {\bibfnamefont {S.}~\bibnamefont
  {Ili\ifmmode~\acute{c}\else \'{c}\fi{}}}\ and\ \bibinfo {author}
  {\bibfnamefont {F.~S.}\ \bibnamefont {Bergeret}},\ }\bibfield  {title}
  {\bibinfo {title} {Theory of the supercurrent diode effect in rashba
  superconductors with arbitrary disorder},\ }\href
  {https://doi.org/10.1103/PhysRevLett.128.177001} {\bibfield  {journal}
  {\bibinfo  {journal} {Phys. Rev. Lett.}\ }\textbf {\bibinfo {volume} {128}},\
  \bibinfo {pages} {177001} (\bibinfo {year} {2022})}\BibitemShut {NoStop}%
\bibitem [{\citenamefont {Scammell}\ \emph {et~al.}(2022)\citenamefont
  {Scammell}, \citenamefont {Li},\ and\ \citenamefont
  {Scheurer}}]{Scammell_2022}%
  \BibitemOpen
  \bibfield  {author} {\bibinfo {author} {\bibfnamefont {H.~D.}\ \bibnamefont
  {Scammell}}, \bibinfo {author} {\bibfnamefont {J.~I.~A.}\ \bibnamefont
  {Li}},\ and\ \bibinfo {author} {\bibfnamefont {M.~S.}\ \bibnamefont
  {Scheurer}},\ }\bibfield  {title} {\bibinfo {title} {Theory of zero-field
  superconducting diode effect in twisted trilayer graphene},\ }\href
  {https://doi.org/10.1088/2053-1583/ac5b16} {\bibfield  {journal} {\bibinfo
  {journal} {2D Materials}\ }\textbf {\bibinfo {volume} {9}},\ \bibinfo {pages}
  {025027} (\bibinfo {year} {2022})}\BibitemShut {NoStop}%
\bibitem [{\citenamefont {Zinkl}\ \emph {et~al.}(2022)\citenamefont {Zinkl},
  \citenamefont {Hamamoto},\ and\ \citenamefont {Sigrist}}]{Zinkl_2022}%
  \BibitemOpen
  \bibfield  {author} {\bibinfo {author} {\bibfnamefont {B.}~\bibnamefont
  {Zinkl}}, \bibinfo {author} {\bibfnamefont {K.}~\bibnamefont {Hamamoto}},\
  and\ \bibinfo {author} {\bibfnamefont {M.}~\bibnamefont {Sigrist}},\
  }\bibfield  {title} {\bibinfo {title} {Symmetry conditions for the
  superconducting diode effect in chiral superconductors},\ }\href
  {https://doi.org/10.1103/PhysRevResearch.4.033167} {\bibfield  {journal}
  {\bibinfo  {journal} {Phys. Rev. Res.}\ }\textbf {\bibinfo {volume} {4}},\
  \bibinfo {pages} {033167} (\bibinfo {year} {2022})}\BibitemShut {NoStop}%
\bibitem [{\citenamefont {He}\ \emph {et~al.}(2023)\citenamefont {He},
  \citenamefont {Tanaka},\ and\ \citenamefont {Nagaosa}}]{He_2023}%
  \BibitemOpen
  \bibfield  {author} {\bibinfo {author} {\bibfnamefont {J.~J.}\ \bibnamefont
  {He}}, \bibinfo {author} {\bibfnamefont {Y.}~\bibnamefont {Tanaka}},\ and\
  \bibinfo {author} {\bibfnamefont {N.}~\bibnamefont {Nagaosa}},\ }\bibfield
  {title} {\bibinfo {title} {The supercurrent diode effect and nonreciprocal
  paraconductivity due to the chiral structure of nanotubes},\ }\href
  {https://doi.org/10.1038/s41467-023-39083-3} {\bibfield  {journal} {\bibinfo
  {journal} {Nature Communications}\ }\textbf {\bibinfo {volume} {14}},\
  \bibinfo {pages} {3330} (\bibinfo {year} {2023})}\BibitemShut {NoStop}%
\bibitem [{\citenamefont {Zhai}\ \emph {et~al.}(2022)\citenamefont {Zhai},
  \citenamefont {Li}, \citenamefont {Wen}, \citenamefont {Wu},\ and\
  \citenamefont {He}}]{Zhai_2022}%
  \BibitemOpen
  \bibfield  {author} {\bibinfo {author} {\bibfnamefont {B.}~\bibnamefont
  {Zhai}}, \bibinfo {author} {\bibfnamefont {B.}~\bibnamefont {Li}}, \bibinfo
  {author} {\bibfnamefont {Y.}~\bibnamefont {Wen}}, \bibinfo {author}
  {\bibfnamefont {F.}~\bibnamefont {Wu}},\ and\ \bibinfo {author}
  {\bibfnamefont {J.}~\bibnamefont {He}},\ }\bibfield  {title} {\bibinfo
  {title} {Prediction of ferroelectric superconductors with reversible
  superconducting diode effect},\ }\href
  {https://doi.org/10.1103/PhysRevB.106.L140505} {\bibfield  {journal}
  {\bibinfo  {journal} {Phys. Rev. B}\ }\textbf {\bibinfo {volume} {106}},\
  \bibinfo {pages} {L140505} (\bibinfo {year} {2022})}\BibitemShut {NoStop}%
\bibitem [{\citenamefont {Jiang}\ \emph {et~al.}(2022)\citenamefont {Jiang},
  \citenamefont {Milo\ifmmode \check{s}\else
  \v{s}\fi{}evi\ifmmode~\acute{c}\else \'{c}\fi{}}, \citenamefont {Wang},
  \citenamefont {Xiao}, \citenamefont {Peeters},\ and\ \citenamefont
  {Chen}}]{Jiang_2022}%
  \BibitemOpen
  \bibfield  {author} {\bibinfo {author} {\bibfnamefont {J.}~\bibnamefont
  {Jiang}}, \bibinfo {author} {\bibfnamefont {M.}~\bibnamefont {Milo\ifmmode
  \check{s}\else \v{s}\fi{}evi\ifmmode~\acute{c}\else \'{c}\fi{}}}, \bibinfo
  {author} {\bibfnamefont {Y.-L.}\ \bibnamefont {Wang}}, \bibinfo {author}
  {\bibfnamefont {Z.-L.}\ \bibnamefont {Xiao}}, \bibinfo {author}
  {\bibfnamefont {F.}~\bibnamefont {Peeters}},\ and\ \bibinfo {author}
  {\bibfnamefont {Q.-H.}\ \bibnamefont {Chen}},\ }\bibfield  {title} {\bibinfo
  {title} {Field-free superconducting diode in a magnetically nanostructured
  superconductor},\ }\href {https://doi.org/10.1103/PhysRevApplied.18.034064}
  {\bibfield  {journal} {\bibinfo  {journal} {Phys. Rev. Appl.}\ }\textbf
  {\bibinfo {volume} {18}},\ \bibinfo {pages} {034064} (\bibinfo {year}
  {2022})}\BibitemShut {NoStop}%
\bibitem [{\citenamefont {de~Picoli}\ \emph {et~al.}(2023)\citenamefont
  {de~Picoli}, \citenamefont {Blood}, \citenamefont {Lyanda-Geller},\ and\
  \citenamefont {V\"ayrynen}}]{Picoli_2023}%
  \BibitemOpen
  \bibfield  {author} {\bibinfo {author} {\bibfnamefont {T.}~\bibnamefont
  {de~Picoli}}, \bibinfo {author} {\bibfnamefont {Z.}~\bibnamefont {Blood}},
  \bibinfo {author} {\bibfnamefont {Y.}~\bibnamefont {Lyanda-Geller}},\ and\
  \bibinfo {author} {\bibfnamefont {J.~I.}\ \bibnamefont {V\"ayrynen}},\
  }\bibfield  {title} {\bibinfo {title} {Superconducting diode effect in
  quasi-one-dimensional systems},\ }\href
  {https://doi.org/10.1103/PhysRevB.107.224518} {\bibfield  {journal} {\bibinfo
   {journal} {Phys. Rev. B}\ }\textbf {\bibinfo {volume} {107}},\ \bibinfo
  {pages} {224518} (\bibinfo {year} {2023})}\BibitemShut {NoStop}%
\bibitem [{\citenamefont {Legg}\ \emph {et~al.}(2022)\citenamefont {Legg},
  \citenamefont {Loss},\ and\ \citenamefont {Klinovaja}}]{Legg_2022}%
  \BibitemOpen
  \bibfield  {author} {\bibinfo {author} {\bibfnamefont {H.~F.}\ \bibnamefont
  {Legg}}, \bibinfo {author} {\bibfnamefont {D.}~\bibnamefont {Loss}},\ and\
  \bibinfo {author} {\bibfnamefont {J.}~\bibnamefont {Klinovaja}},\ }\bibfield
  {title} {\bibinfo {title} {Superconducting diode effect due to magnetochiral
  anisotropy in topological insulators and rashba nanowires},\ }\href
  {https://doi.org/10.1103/PhysRevB.106.104501} {\bibfield  {journal} {\bibinfo
   {journal} {Phys. Rev. B}\ }\textbf {\bibinfo {volume} {106}},\ \bibinfo
  {pages} {104501} (\bibinfo {year} {2022})}\BibitemShut {NoStop}%
\bibitem [{\citenamefont {Banerjee}\ and\ \citenamefont
  {Scheurer}(2024)}]{Sayan_2024}%
  \BibitemOpen
  \bibfield  {author} {\bibinfo {author} {\bibfnamefont {S.}~\bibnamefont
  {Banerjee}}\ and\ \bibinfo {author} {\bibfnamefont {M.~S.}\ \bibnamefont
  {Scheurer}},\ }\bibfield  {title} {\bibinfo {title} {Enhanced superconducting
  diode effect due to coexisting phases},\ }\href
  {https://doi.org/10.1103/PhysRevLett.132.046003} {\bibfield  {journal}
  {\bibinfo  {journal} {Phys. Rev. Lett.}\ }\textbf {\bibinfo {volume} {132}},\
  \bibinfo {pages} {046003} (\bibinfo {year} {2024})}\BibitemShut {NoStop}%
\bibitem [{\citenamefont {Bhowmik}\ \emph {et~al.}(2025)\citenamefont
  {Bhowmik}, \citenamefont {Samanta}, \citenamefont {Nandy}, \citenamefont
  {Saha},\ and\ \citenamefont {Ghosh}}]{bhowmik_2025}%
  \BibitemOpen
  \bibfield  {author} {\bibinfo {author} {\bibfnamefont {S.}~\bibnamefont
  {Bhowmik}}, \bibinfo {author} {\bibfnamefont {D.}~\bibnamefont {Samanta}},
  \bibinfo {author} {\bibfnamefont {A.~K.}\ \bibnamefont {Nandy}}, \bibinfo
  {author} {\bibfnamefont {A.}~\bibnamefont {Saha}},\ and\ \bibinfo {author}
  {\bibfnamefont {S.~K.}\ \bibnamefont {Ghosh}},\ }\bibfield  {title} {\bibinfo
  {title} {Optimizing one dimensional superconducting diodes: interplay of
  rashba spin-orbit coupling and magnetic fields},\ }\href
  {https://doi.org/10.1038/s42005-025-02044-x} {\bibfield  {journal} {\bibinfo
  {journal} {Communications Physics}\ }\textbf {\bibinfo {volume} {8}},\
  \bibinfo {pages} {260} (\bibinfo {year} {2025})}\BibitemShut {NoStop}%
\bibitem [{\citenamefont {Samanta}\ and\ \citenamefont
  {Ghosh}(2025)}]{dibyendu_2025}%
  \BibitemOpen
  \bibfield  {author} {\bibinfo {author} {\bibfnamefont {D.}~\bibnamefont
  {Samanta}}\ and\ \bibinfo {author} {\bibfnamefont {S.~K.}\ \bibnamefont
  {Ghosh}},\ }\bibfield  {title} {\bibinfo {title} {Field-free superconducting
  diode effect and topological fulde-ferrell-larkin-ovchinnikov
  superconductivity in altermagnetic shiba chains},\ }\href
  {https://arxiv.org/abs/2507.21446} {\bibfield  {journal} {\bibinfo  {journal}
  {arXiv:2507.21446}\ } (\bibinfo {year} {2025})}\BibitemShut {NoStop}%
\bibitem [{\citenamefont {Hou}\ \emph {et~al.}(2023)\citenamefont {Hou},
  \citenamefont {Nichele}, \citenamefont {Chi}, \citenamefont {Lodesani},
  \citenamefont {Wu}, \citenamefont {Ritter}, \citenamefont {Haxell},
  \citenamefont {Davydova}, \citenamefont {Ili\ifmmode~\acute{c}\else
  \'{c}\fi{}}, \citenamefont {Glezakou-Elbert}, \citenamefont {Varambally},
  \citenamefont {Bergeret}, \citenamefont {Kamra}, \citenamefont {Fu},
  \citenamefont {Lee},\ and\ \citenamefont {Moodera}}]{Hou_2023}%
  \BibitemOpen
  \bibfield  {author} {\bibinfo {author} {\bibfnamefont {Y.}~\bibnamefont
  {Hou}}, \bibinfo {author} {\bibfnamefont {F.}~\bibnamefont {Nichele}},
  \bibinfo {author} {\bibfnamefont {H.}~\bibnamefont {Chi}}, \bibinfo {author}
  {\bibfnamefont {A.}~\bibnamefont {Lodesani}}, \bibinfo {author}
  {\bibfnamefont {Y.}~\bibnamefont {Wu}}, \bibinfo {author} {\bibfnamefont
  {M.~F.}\ \bibnamefont {Ritter}}, \bibinfo {author} {\bibfnamefont {D.~Z.}\
  \bibnamefont {Haxell}}, \bibinfo {author} {\bibfnamefont {M.}~\bibnamefont
  {Davydova}}, \bibinfo {author} {\bibfnamefont {S.}~\bibnamefont
  {Ili\ifmmode~\acute{c}\else \'{c}\fi{}}}, \bibinfo {author} {\bibfnamefont
  {O.}~\bibnamefont {Glezakou-Elbert}}, \bibinfo {author} {\bibfnamefont
  {A.}~\bibnamefont {Varambally}}, \bibinfo {author} {\bibfnamefont {F.~S.}\
  \bibnamefont {Bergeret}}, \bibinfo {author} {\bibfnamefont {A.}~\bibnamefont
  {Kamra}}, \bibinfo {author} {\bibfnamefont {L.}~\bibnamefont {Fu}}, \bibinfo
  {author} {\bibfnamefont {P.~A.}\ \bibnamefont {Lee}},\ and\ \bibinfo {author}
  {\bibfnamefont {J.~S.}\ \bibnamefont {Moodera}},\ }\bibfield  {title}
  {\bibinfo {title} {Ubiquitous superconducting diode effect in superconductor
  thin films},\ }\href {https://doi.org/10.1103/PhysRevLett.131.027001}
  {\bibfield  {journal} {\bibinfo  {journal} {Phys. Rev. Lett.}\ }\textbf
  {\bibinfo {volume} {131}},\ \bibinfo {pages} {027001} (\bibinfo {year}
  {2023})}\BibitemShut {NoStop}%
\bibitem [{\citenamefont {Gupta}\ \emph {et~al.}(2023)\citenamefont {Gupta},
  \citenamefont {Graziano}, \citenamefont {Pendharkar}, \citenamefont {Dong},
  \citenamefont {Dempsey}, \citenamefont {Palmstr{\o}m},\ and\ \citenamefont
  {Pribiag}}]{gupta_2023}%
  \BibitemOpen
  \bibfield  {author} {\bibinfo {author} {\bibfnamefont {M.}~\bibnamefont
  {Gupta}}, \bibinfo {author} {\bibfnamefont {G.~V.}\ \bibnamefont {Graziano}},
  \bibinfo {author} {\bibfnamefont {M.}~\bibnamefont {Pendharkar}}, \bibinfo
  {author} {\bibfnamefont {J.~T.}\ \bibnamefont {Dong}}, \bibinfo {author}
  {\bibfnamefont {C.~P.}\ \bibnamefont {Dempsey}}, \bibinfo {author}
  {\bibfnamefont {C.}~\bibnamefont {Palmstr{\o}m}},\ and\ \bibinfo {author}
  {\bibfnamefont {V.~S.}\ \bibnamefont {Pribiag}},\ }\bibfield  {title}
  {\bibinfo {title} {Gate-tunable superconducting diode effect in a
  three-terminal josephson device},\ }\href
  {https://doi.org/10.1038/s41467-023-38856-0} {\bibfield  {journal} {\bibinfo
  {journal} {Nature communications}\ }\textbf {\bibinfo {volume} {14}},\
  \bibinfo {pages} {3078} (\bibinfo {year} {2023})}\BibitemShut {NoStop}%
\bibitem [{\citenamefont {Banerjee}\ \emph {et~al.}(2023)\citenamefont
  {Banerjee}, \citenamefont {Geier}, \citenamefont {Rahman}, \citenamefont
  {Thomas}, \citenamefont {Wang}, \citenamefont {Manfra}, \citenamefont
  {Flensberg},\ and\ \citenamefont {Marcus}}]{Abhishek_2023}%
  \BibitemOpen
  \bibfield  {author} {\bibinfo {author} {\bibfnamefont {A.}~\bibnamefont
  {Banerjee}}, \bibinfo {author} {\bibfnamefont {M.}~\bibnamefont {Geier}},
  \bibinfo {author} {\bibfnamefont {M.~A.}\ \bibnamefont {Rahman}}, \bibinfo
  {author} {\bibfnamefont {C.}~\bibnamefont {Thomas}}, \bibinfo {author}
  {\bibfnamefont {T.}~\bibnamefont {Wang}}, \bibinfo {author} {\bibfnamefont
  {M.~J.}\ \bibnamefont {Manfra}}, \bibinfo {author} {\bibfnamefont
  {K.}~\bibnamefont {Flensberg}},\ and\ \bibinfo {author} {\bibfnamefont
  {C.~M.}\ \bibnamefont {Marcus}},\ }\bibfield  {title} {\bibinfo {title}
  {Phase asymmetry of andreev spectra from cooper-pair momentum},\ }\href
  {https://doi.org/10.1103/PhysRevLett.131.196301} {\bibfield  {journal}
  {\bibinfo  {journal} {Phys. Rev. Lett.}\ }\textbf {\bibinfo {volume} {131}},\
  \bibinfo {pages} {196301} (\bibinfo {year} {2023})}\BibitemShut {NoStop}%
\bibitem [{\citenamefont {Narita}\ \emph {et~al.}(2022)\citenamefont {Narita},
  \citenamefont {Ishizuka}, \citenamefont {Kawarazaki}, \citenamefont {Kan},
  \citenamefont {Shiota}, \citenamefont {Moriyama}, \citenamefont {Shimakawa},
  \citenamefont {Ognev}, \citenamefont {Samardak}, \citenamefont {Yanase} \emph
  {et~al.}}]{narita_2022}%
  \BibitemOpen
  \bibfield  {author} {\bibinfo {author} {\bibfnamefont {H.}~\bibnamefont
  {Narita}}, \bibinfo {author} {\bibfnamefont {J.}~\bibnamefont {Ishizuka}},
  \bibinfo {author} {\bibfnamefont {R.}~\bibnamefont {Kawarazaki}}, \bibinfo
  {author} {\bibfnamefont {D.}~\bibnamefont {Kan}}, \bibinfo {author}
  {\bibfnamefont {Y.}~\bibnamefont {Shiota}}, \bibinfo {author} {\bibfnamefont
  {T.}~\bibnamefont {Moriyama}}, \bibinfo {author} {\bibfnamefont
  {Y.}~\bibnamefont {Shimakawa}}, \bibinfo {author} {\bibfnamefont {A.~V.}\
  \bibnamefont {Ognev}}, \bibinfo {author} {\bibfnamefont {A.~S.}\ \bibnamefont
  {Samardak}}, \bibinfo {author} {\bibfnamefont {Y.}~\bibnamefont {Yanase}},
  \emph {et~al.},\ }\bibfield  {title} {\bibinfo {title} {Field-free
  superconducting diode effect in noncentrosymmetric superconductor/ferromagnet
  multilayers},\ }\href {https://doi.org/10.1038/s41565-022-01159-4} {\bibfield
   {journal} {\bibinfo  {journal} {Nature Nanotechnology}\ }\textbf {\bibinfo
  {volume} {17}},\ \bibinfo {pages} {823} (\bibinfo {year} {2022})}\BibitemShut
  {NoStop}%
\bibitem [{\citenamefont {Gutfreund}\ \emph {et~al.}(2023)\citenamefont
  {Gutfreund}, \citenamefont {Matsuki}, \citenamefont {Plastovets},
  \citenamefont {Noah}, \citenamefont {Gorzawski}, \citenamefont {Fridman},
  \citenamefont {Yang}, \citenamefont {Buzdin}, \citenamefont {Millo},
  \citenamefont {Robinson} \emph {et~al.}}]{gutfreund_2023}%
  \BibitemOpen
  \bibfield  {author} {\bibinfo {author} {\bibfnamefont {A.}~\bibnamefont
  {Gutfreund}}, \bibinfo {author} {\bibfnamefont {H.}~\bibnamefont {Matsuki}},
  \bibinfo {author} {\bibfnamefont {V.}~\bibnamefont {Plastovets}}, \bibinfo
  {author} {\bibfnamefont {A.}~\bibnamefont {Noah}}, \bibinfo {author}
  {\bibfnamefont {L.}~\bibnamefont {Gorzawski}}, \bibinfo {author}
  {\bibfnamefont {N.}~\bibnamefont {Fridman}}, \bibinfo {author} {\bibfnamefont
  {G.}~\bibnamefont {Yang}}, \bibinfo {author} {\bibfnamefont {A.}~\bibnamefont
  {Buzdin}}, \bibinfo {author} {\bibfnamefont {O.}~\bibnamefont {Millo}},
  \bibinfo {author} {\bibfnamefont {J.~W.}\ \bibnamefont {Robinson}}, \emph
  {et~al.},\ }\bibfield  {title} {\bibinfo {title} {Direct observation of a
  superconducting vortex diode},\ }\href
  {https://doi.org/10.1038/s41467-023-37294-2} {\bibfield  {journal} {\bibinfo
  {journal} {Nature Communications}\ }\textbf {\bibinfo {volume} {14}},\
  \bibinfo {pages} {1630} (\bibinfo {year} {2023})}\BibitemShut {NoStop}%
\bibitem [{\citenamefont {Chen}\ \emph {et~al.}(2025)\citenamefont {Chen},
  \citenamefont {Scheurer},\ and\ \citenamefont {Schrade}}]{chen_2025}%
  \BibitemOpen
  \bibfield  {author} {\bibinfo {author} {\bibfnamefont {Y.}~\bibnamefont
  {Chen}}, \bibinfo {author} {\bibfnamefont {M.~S.}\ \bibnamefont {Scheurer}},\
  and\ \bibinfo {author} {\bibfnamefont {C.}~\bibnamefont {Schrade}},\
  }\bibfield  {title} {\bibinfo {title} {Intrinsic superconducting diode effect
  and nonreciprocal superconductivity in rhombohedral graphene multilayers},\
  }\href@noop {} {\bibfield  {journal} {\bibinfo  {journal} {arXiv preprint
  arXiv:2503.16391}\ } (\bibinfo {year} {2025})}\BibitemShut {NoStop}%
\bibitem [{\citenamefont {Choy}\ \emph {et~al.}(2011)\citenamefont {Choy},
  \citenamefont {Edge}, \citenamefont {Akhmerov},\ and\ \citenamefont
  {Beenakker}}]{Choy_2011}%
  \BibitemOpen
  \bibfield  {author} {\bibinfo {author} {\bibfnamefont {T.-P.}\ \bibnamefont
  {Choy}}, \bibinfo {author} {\bibfnamefont {J.~M.}\ \bibnamefont {Edge}},
  \bibinfo {author} {\bibfnamefont {A.~R.}\ \bibnamefont {Akhmerov}},\ and\
  \bibinfo {author} {\bibfnamefont {C.~W.~J.}\ \bibnamefont {Beenakker}},\
  }\bibfield  {title} {\bibinfo {title} {Majorana fermions emerging from
  magnetic nanoparticles on a superconductor without spin-orbit coupling},\
  }\href {https://doi.org/10.1103/PhysRevB.84.195442} {\bibfield  {journal}
  {\bibinfo  {journal} {Phys. Rev. B}\ }\textbf {\bibinfo {volume} {84}},\
  \bibinfo {pages} {195442} (\bibinfo {year} {2011})}\BibitemShut {NoStop}%
\bibitem [{\citenamefont {Nadj-Perge}\ \emph {et~al.}(2013)\citenamefont
  {Nadj-Perge}, \citenamefont {Drozdov}, \citenamefont {Bernevig},\ and\
  \citenamefont {Yazdani}}]{Nadj-Perge_2013}%
  \BibitemOpen
  \bibfield  {author} {\bibinfo {author} {\bibfnamefont {S.}~\bibnamefont
  {Nadj-Perge}}, \bibinfo {author} {\bibfnamefont {I.~K.}\ \bibnamefont
  {Drozdov}}, \bibinfo {author} {\bibfnamefont {B.~A.}\ \bibnamefont
  {Bernevig}},\ and\ \bibinfo {author} {\bibfnamefont {A.}~\bibnamefont
  {Yazdani}},\ }\bibfield  {title} {\bibinfo {title} {Proposal for realizing
  majorana fermions in chains of magnetic atoms on a superconductor},\ }\href
  {https://doi.org/10.1103/PhysRevB.88.020407} {\bibfield  {journal} {\bibinfo
  {journal} {Phys. Rev. B}\ }\textbf {\bibinfo {volume} {88}},\ \bibinfo
  {pages} {020407} (\bibinfo {year} {2013})}\BibitemShut {NoStop}%
\bibitem [{\citenamefont {Pientka}\ \emph {et~al.}(2013)\citenamefont
  {Pientka}, \citenamefont {Glazman},\ and\ \citenamefont {von
  Oppen}}]{Pientka_2013}%
  \BibitemOpen
  \bibfield  {author} {\bibinfo {author} {\bibfnamefont {F.}~\bibnamefont
  {Pientka}}, \bibinfo {author} {\bibfnamefont {L.~I.}\ \bibnamefont
  {Glazman}},\ and\ \bibinfo {author} {\bibfnamefont {F.}~\bibnamefont {von
  Oppen}},\ }\bibfield  {title} {\bibinfo {title} {Topological superconducting
  phase in helical shiba chains},\ }\href
  {https://doi.org/10.1103/PhysRevB.88.155420} {\bibfield  {journal} {\bibinfo
  {journal} {Phys. Rev. B}\ }\textbf {\bibinfo {volume} {88}},\ \bibinfo
  {pages} {155420} (\bibinfo {year} {2013})}\BibitemShut {NoStop}%
\bibitem [{\citenamefont {Vazifeh}\ and\ \citenamefont
  {Franz}(2013)}]{Vazifeh_2013}%
  \BibitemOpen
  \bibfield  {author} {\bibinfo {author} {\bibfnamefont {M.~M.}\ \bibnamefont
  {Vazifeh}}\ and\ \bibinfo {author} {\bibfnamefont {M.}~\bibnamefont
  {Franz}},\ }\bibfield  {title} {\bibinfo {title} {Self-organized topological
  state with majorana fermions},\ }\href
  {https://doi.org/10.1103/PhysRevLett.111.206802} {\bibfield  {journal}
  {\bibinfo  {journal} {Phys. Rev. Lett.}\ }\textbf {\bibinfo {volume} {111}},\
  \bibinfo {pages} {206802} (\bibinfo {year} {2013})}\BibitemShut {NoStop}%
\bibitem [{\citenamefont {Heimes}\ \emph {et~al.}(2014)\citenamefont {Heimes},
  \citenamefont {Kotetes},\ and\ \citenamefont {Sch\"on}}]{Heimes_2014}%
  \BibitemOpen
  \bibfield  {author} {\bibinfo {author} {\bibfnamefont {A.}~\bibnamefont
  {Heimes}}, \bibinfo {author} {\bibfnamefont {P.}~\bibnamefont {Kotetes}},\
  and\ \bibinfo {author} {\bibfnamefont {G.}~\bibnamefont {Sch\"on}},\
  }\bibfield  {title} {\bibinfo {title} {Majorana fermions from shiba states in
  an antiferromagnetic chain on top of a superconductor},\ }\href
  {https://doi.org/10.1103/PhysRevB.90.060507} {\bibfield  {journal} {\bibinfo
  {journal} {Phys. Rev. B}\ }\textbf {\bibinfo {volume} {90}},\ \bibinfo
  {pages} {060507} (\bibinfo {year} {2014})}\BibitemShut {NoStop}%
\bibitem [{\citenamefont {Bhowmik}\ and\ \citenamefont
  {Saha}(2025)}]{Sayak_2025}%
  \BibitemOpen
  \bibfield  {author} {\bibinfo {author} {\bibfnamefont {S.}~\bibnamefont
  {Bhowmik}}\ and\ \bibinfo {author} {\bibfnamefont {A.}~\bibnamefont {Saha}},\
  }\bibfield  {title} {\bibinfo {title} {Topological majorana zero modes and
  the superconducting diode effect driven by fulde-ferrell-larkin-ovchinnikov
  pairing in a helical shiba chain},\ }\href
  {https://doi.org/10.1103/PhysRevB.111.L161402} {\bibfield  {journal}
  {\bibinfo  {journal} {Phys. Rev. B}\ }\textbf {\bibinfo {volume} {111}},\
  \bibinfo {pages} {L161402} (\bibinfo {year} {2025})}\BibitemShut {NoStop}%
\bibitem [{\citenamefont {Fulde}\ and\ \citenamefont
  {Ferrell}(1964)}]{Fulde_1964}%
  \BibitemOpen
  \bibfield  {author} {\bibinfo {author} {\bibfnamefont {P.}~\bibnamefont
  {Fulde}}\ and\ \bibinfo {author} {\bibfnamefont {R.~A.}\ \bibnamefont
  {Ferrell}},\ }\bibfield  {title} {\bibinfo {title} {Superconductivity in a
  strong spin-exchange field},\ }\href
  {https://doi.org/10.1103/PhysRev.135.A550} {\bibfield  {journal} {\bibinfo
  {journal} {Phys. Rev.}\ }\textbf {\bibinfo {volume} {135}},\ \bibinfo {pages}
  {A550} (\bibinfo {year} {1964})}\BibitemShut {NoStop}%
\bibitem [{\citenamefont {Larkin}\ and\ \citenamefont
  {Ovchinnikov}(1965)}]{larkin_1965}%
  \BibitemOpen
  \bibfield  {author} {\bibinfo {author} {\bibfnamefont {A.}~\bibnamefont
  {Larkin}}\ and\ \bibinfo {author} {\bibfnamefont {Y.~N.}\ \bibnamefont
  {Ovchinnikov}},\ }\bibfield  {title} {\bibinfo {title} {Nonuniform state of
  superconductors},\ }\href@noop {} {\bibfield  {journal} {\bibinfo  {journal}
  {Soviet Physics-JETP}\ }\textbf {\bibinfo {volume} {20}},\ \bibinfo {pages}
  {762} (\bibinfo {year} {1965})}\BibitemShut {NoStop}%
\bibitem [{\citenamefont {Lo~Conte}\ \emph {et~al.}(2025)\citenamefont
  {Lo~Conte}, \citenamefont {Wiebe}, \citenamefont {Rachel}, \citenamefont
  {Morr},\ and\ \citenamefont {Wiesendanger}}]{Weis}%
  \BibitemOpen
  \bibfield  {author} {\bibinfo {author} {\bibfnamefont {R.}~\bibnamefont
  {Lo~Conte}}, \bibinfo {author} {\bibfnamefont {J.}~\bibnamefont {Wiebe}},
  \bibinfo {author} {\bibfnamefont {S.}~\bibnamefont {Rachel}}, \bibinfo
  {author} {\bibfnamefont {D.~K.}\ \bibnamefont {Morr}},\ and\ \bibinfo
  {author} {\bibfnamefont {R.}~\bibnamefont {Wiesendanger}},\ }\bibfield
  {title} {\bibinfo {title} {Magnet-superconductor hybrid quantum systems: a
  materials platform for topological superconductivity},\ }\href
  {https://doi.org/10.1007/s40766-024-00060-1} {\bibfield  {journal} {\bibinfo
  {journal} {La Rivista del Nuovo Cimento}\ ,\ \bibinfo {pages} {1}} (\bibinfo
  {year} {2025})}\BibitemShut {NoStop}%
\bibitem [{\citenamefont {Hess}\ \emph {et~al.}(2022)\citenamefont {Hess},
  \citenamefont {Legg}, \citenamefont {Loss},\ and\ \citenamefont
  {Klinovaja}}]{Richard_2022}%
  \BibitemOpen
  \bibfield  {author} {\bibinfo {author} {\bibfnamefont {R.}~\bibnamefont
  {Hess}}, \bibinfo {author} {\bibfnamefont {H.~F.}\ \bibnamefont {Legg}},
  \bibinfo {author} {\bibfnamefont {D.}~\bibnamefont {Loss}},\ and\ \bibinfo
  {author} {\bibfnamefont {J.}~\bibnamefont {Klinovaja}},\ }\bibfield  {title}
  {\bibinfo {title} {Prevalence of trivial zero-energy subgap states in
  nonuniform helical spin chains on the surface of superconductors},\ }\href
  {https://doi.org/10.1103/PhysRevB.106.104503} {\bibfield  {journal} {\bibinfo
   {journal} {Phys. Rev. B}\ }\textbf {\bibinfo {volume} {106}},\ \bibinfo
  {pages} {104503} (\bibinfo {year} {2022})}\BibitemShut {NoStop}%
\bibitem [{\citenamefont {Zhu}(2016)}]{Zhu_2016}%
  \BibitemOpen
  \bibfield  {author} {\bibinfo {author} {\bibfnamefont {J.-X.}\ \bibnamefont
  {Zhu}},\ }\href@noop {} {\emph {\bibinfo {title} {Bogoliubov-de Gennes method
  and its applications}}},\ Vol.\ \bibinfo {volume} {924}\ (\bibinfo
  {publisher} {Springer},\ \bibinfo {year} {2016})\BibitemShut {NoStop}%
\bibitem [{\citenamefont {Br{\"u}ning}\ \emph {et~al.}(2024)\citenamefont
  {Br{\"u}ning}, \citenamefont {Bedow}, \citenamefont {Conte}, \citenamefont
  {von Bergmann}, \citenamefont {Morr}, \citenamefont {Wiesendanger} \emph
  {et~al.}}]{Roland}%
  \BibitemOpen
  \bibfield  {author} {\bibinfo {author} {\bibfnamefont {R.}~\bibnamefont
  {Br{\"u}ning}}, \bibinfo {author} {\bibfnamefont {J.}~\bibnamefont {Bedow}},
  \bibinfo {author} {\bibfnamefont {R.~L.}\ \bibnamefont {Conte}}, \bibinfo
  {author} {\bibfnamefont {K.}~\bibnamefont {von Bergmann}}, \bibinfo {author}
  {\bibfnamefont {D.}~\bibnamefont {Morr}}, \bibinfo {author} {\bibfnamefont
  {R.}~\bibnamefont {Wiesendanger}}, \emph {et~al.},\ }\bibfield  {title}
  {\bibinfo {title} {The non-collinear path to topological superconductivity},\
  }\href@noop {} {\bibfield  {journal} {\bibinfo  {journal} {arXiv preprint
  arXiv:2405.14673}\ } (\bibinfo {year} {2024})}\BibitemShut {NoStop}%
\bibitem [{\citenamefont {R\'ozsa}\ \emph {et~al.}(2015)\citenamefont
  {R\'ozsa}, \citenamefont {Udvardi}, \citenamefont {Szunyogh},\ and\
  \citenamefont {Szab\'o}}]{DFT}%
  \BibitemOpen
  \bibfield  {author} {\bibinfo {author} {\bibfnamefont {L.}~\bibnamefont
  {R\'ozsa}}, \bibinfo {author} {\bibfnamefont {L.}~\bibnamefont {Udvardi}},
  \bibinfo {author} {\bibfnamefont {L.}~\bibnamefont {Szunyogh}},\ and\
  \bibinfo {author} {\bibfnamefont {I.~A.}\ \bibnamefont {Szab\'o}},\
  }\bibfield  {title} {\bibinfo {title} {Magnetic phase diagram of an fe
  monolayer on w(110) and ta(110) surfaces based on ab initio calculations},\
  }\href {https://doi.org/10.1103/PhysRevB.91.144424} {\bibfield  {journal}
  {\bibinfo  {journal} {Phys. Rev. B}\ }\textbf {\bibinfo {volume} {91}},\
  \bibinfo {pages} {144424} (\bibinfo {year} {2015})}\BibitemShut {NoStop}%
\bibitem [{\citenamefont {Chatterjee}\ \emph {et~al.}(2024)\citenamefont
  {Chatterjee}, \citenamefont {Banik}, \citenamefont {Bera}, \citenamefont
  {Ghosh}, \citenamefont {Pradhan}, \citenamefont {Saha},\ and\ \citenamefont
  {Nandy}}]{Pritam_2024}%
  \BibitemOpen
  \bibfield  {author} {\bibinfo {author} {\bibfnamefont {P.}~\bibnamefont
  {Chatterjee}}, \bibinfo {author} {\bibfnamefont {S.}~\bibnamefont {Banik}},
  \bibinfo {author} {\bibfnamefont {S.}~\bibnamefont {Bera}}, \bibinfo {author}
  {\bibfnamefont {A.~K.}\ \bibnamefont {Ghosh}}, \bibinfo {author}
  {\bibfnamefont {S.}~\bibnamefont {Pradhan}}, \bibinfo {author} {\bibfnamefont
  {A.}~\bibnamefont {Saha}},\ and\ \bibinfo {author} {\bibfnamefont {A.~K.}\
  \bibnamefont {Nandy}},\ }\bibfield  {title} {\bibinfo {title} {Topological
  superconductivity by engineering noncollinear magnetism in
  magnet/superconductor heterostructures: A realistic prescription for the
  two-dimensional kitaev model},\ }\href
  {https://doi.org/10.1103/PhysRevB.109.L121301} {\bibfield  {journal}
  {\bibinfo  {journal} {Phys. Rev. B}\ }\textbf {\bibinfo {volume} {109}},\
  \bibinfo {pages} {L121301} (\bibinfo {year} {2024})}\BibitemShut {NoStop}%
\end{thebibliography}%

\clearpage
\onecolumngrid 

\section*{Supplementary material to ``Field-free superconducting diode effect in two-dimensional Shiba lattices''}

\setcounter{section}{0} 
\renewcommand{\thesection}{S\arabic{section}} 


This supplementary text provides an in-depth analysis that complements the main manuscript. We first investigate the optimization of the superconducting diode effect (SDE) using the lattice-regularized model of a 2D Shiba lattice with conical spin texture. Next, we examine the asymmetry in the Bogoliubov quasiparticle spectra to gain microscopic insight into the origin of the SDE. Finally, we explore alternative spin textures, including planar helical and antiferromagnetic spin spiral, deposited on the surface of a 3D $s$-wave superconductor in the presence 
of an external magnetic field.

\section{Optimization of diode efficiency for lattice regularized model}

\begin{figure}[h]
    \centering 
    \includegraphics[width=\columnwidth]{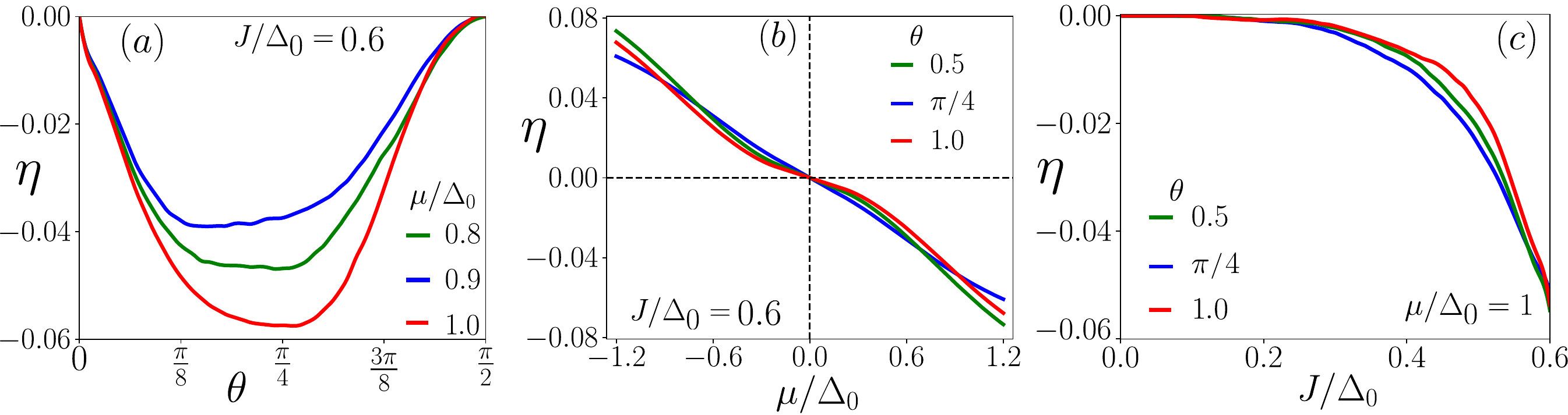}
    \caption{\textbf{Efficiency optimization for the lattice regularized model:} (a) Diode efficiency $\eta$ is shown as a function of $\theta$ for different values of the effective chemical potential $\mu'$, with $J/\Delta_0 = 0.6$. The effective chemical potential is defined as $\mu' = \mu - 4t - (t/4)(g_x^2 + g_y^2)$. (b) Dependence of $\eta$ on $\mu'$ for the optimal choice of $\theta$, with $J/\Delta_0 = 0.6$. 
    (c) $\eta$ is depicted as a function of $J$ for optimized values of both $\theta$ and $\mu'/\Delta_0 = 1.0$. Other model parameters used are: $(t/\Delta_0, g_x, g_y, \alpha, \beta^{-1}/\Delta_0) = (1.0, \pi/2, \pi/2, \pi/4, 0.1)$.}   
    \label{fig:OPT}
\end{figure}

We begin by analyzing the optimization of diode efficiency $\eta$ through systematic variation of key system parameters, aiming to identify conditions that maximize $\eta$ and to understand its dependence on the underlying characteristics of spin texture [see Fig.~\ref{fig:OPT}]. In Fig.~\ref{fig:OPT}(a), $\eta$ is shown as a function of the spin texture cone angle $\theta$ for several values of the chemical potential $\mu$. The efficiency is finite only within the range $0 < \theta < \pi/2$ and vanishes for both trivial ($\theta = 0$) and planar ($\theta = \pi/2$) spin configurations, which do not support a finite FFLO ground-state Cooper pair momentum $q_0$ [see the main text for discussion]. Fig.~\ref{fig:OPT}(b) shows the variation of $\eta$ with $\mu$ for fixed values of $\theta$, revealing a characteristic behavior: $\eta$ increases from zero at $\mu = 0$, reaches a sharp peak, and then decreases rapidly. Moreover, $\eta$ changes sign upon reversing the polarity of $\mu$, reflecting the system’s particle-hole asymmetry. The value of $\mu$ at which $\eta$ peaks depends sensitively on other parameters, such as the exchange coupling $J$ and the cone angle $\theta$. In Fig.~\ref{fig:OPT}(c), the corresponding behavior of $\eta$ is depicted as a function 
of the exchange coupling $J$. Here, $\eta$ monotonically decreases as we increase $J$. Throughout Fig.~\ref{fig:OPT}, the azimuthal angle is fixed at $\alpha = \pi/4$, where the diode efficiency is found to be most pronounced for the chosen conical spin texture configuration.

\section{SDE analysis from momentum space Hamiltonian }

\begin{figure}[b]
\centering 
\includegraphics[width=\columnwidth]{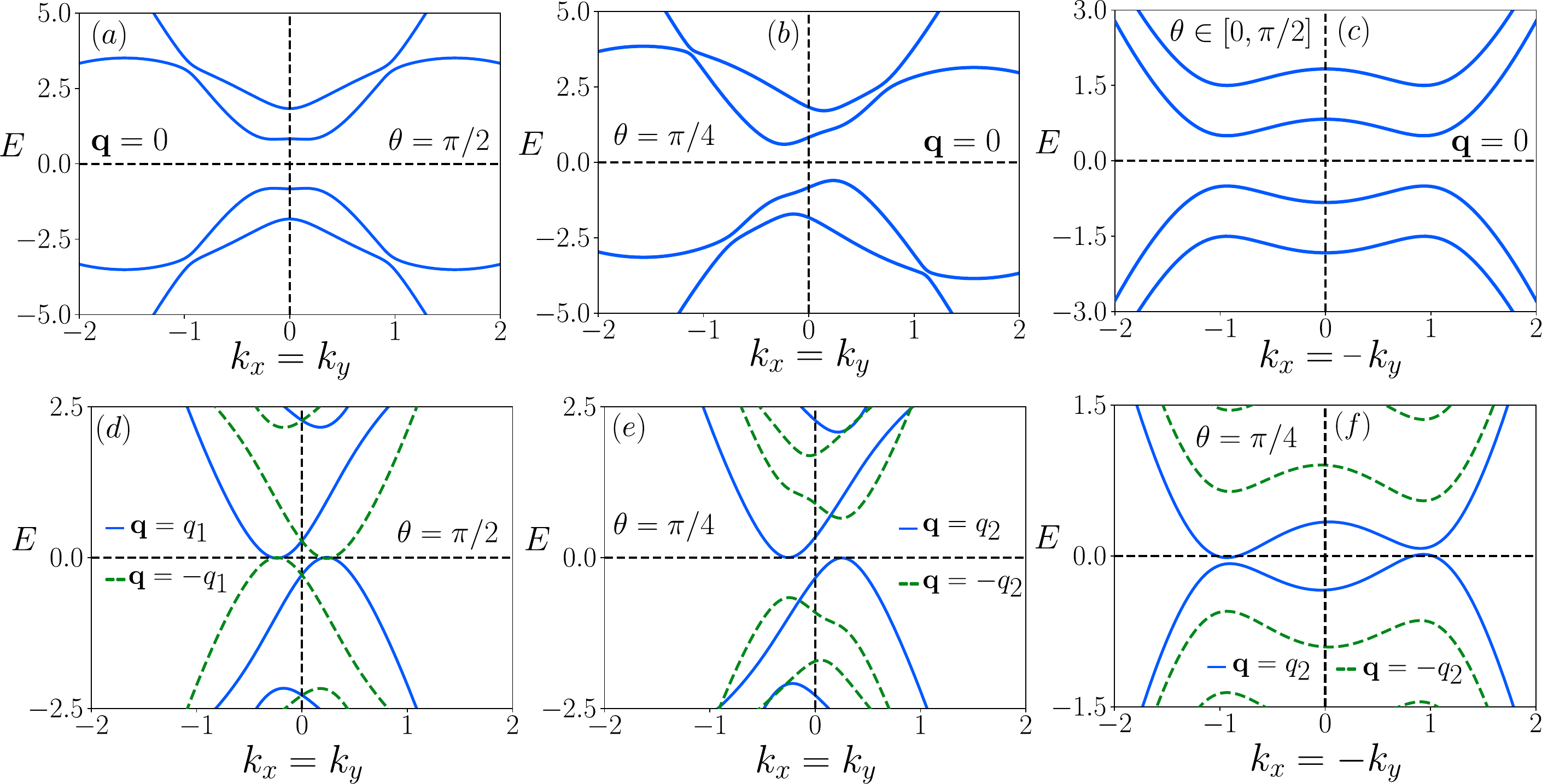}
\caption{\textbf{BDG band structure and properties:} The BDG  band spectrum of the continuum model for a conventional $s$-wave superconducting state with $q = 0$ is shown in panels (a)-(c) considering different spin textures. (a) Depicts the spectrum along $k_x = k_y$ line for a planar spin texture with $\theta = \pi/2$  and (b) for the spin texture in cone phase with $\theta = \pi/4$. In panel (c), the spectrum is shown for cone-phase configurations with cone angles $\theta \in (0, \pi/2)$, evaluated along the $k_x = -k_y$ line. Panels (d)-(f) displays the BDG band spectrum for an FFLO superconducting state with a finite Cooper pair momentum $\mathbf{q}$ for both positive (blue solid line) and negative (dashed green line) $\mathbf{q} $ values.  Panel (d) and (e) show the BdG spectra along the $k_x = k_y$ line considering  planar ($\theta = \pi/2$)  and cone-phase ($\theta = \pi/4$) spin texture, respectively. (f) Illustrates the spectrum along the $k_x = -k_y$ line for the conical spin texture with $\theta = \pi/4$. The system parameters are chosen as $J/\Delta_0=0.5, \mu/\Delta_0=1, g_x=g_y=\pi/2, t/\Delta_0=0.5, q_1=(0.65,0.45),q_2=(0.4,0.3)$.}   
\label{fig:band}
\end{figure}

To investigate the origin of symmetry requirement for the diode current, we analyze the Bogoliubov quasiparticle spectra of the Hamiltonian

\begin{align}
	\mathcal{H}(\mathbf{k}, \mathbf{q}) &=
	\begin{bmatrix}
		h_{\mathbf{k}+\mathbf{q}/2} & \Delta \\
		\Delta & -h^*(-\mathbf{k} + \mathbf{q}/2)
	\end{bmatrix}, \nonumber \\
	h_{\mathbf{k}} &= \epsilon_{\mathbf{k},\tilde{\mathbf{g}}} + \frac{1}{2}(\tilde{\mathbf{g}} \cdot \mathbf{k}) \sigma_z + J \sin(\theta) \sigma_x + J \cos(\theta) \sigma_z \ .
	\label{eq:k_space_hamiltonian}
\end{align}

\noin Here, $\eps_{\mathbf{k},\mathbf{g}}= \frac{1}{2} (\mathbf{k}^2 + \mathbf{\tilde{g}}^2) - \mu$, with $\mathbf{\tilde{g}}=\mathbf{g}/2$. 
We first examine the Bogoliubov quasiparticle spectra with the Hamiltonian for $\mathbf{q}=0$. 
In this case, the spectrum is symmetric for the planar spin texture [see \fig{fig:band}(a)], but exhibits asymmetry for the conical configuration [\fig{fig:band}(b)]. Notably, this asymmetry vanishes only along the $k_x = -k_y$ direction [\fig{fig:band}(c)], where inversion symmetry is preserved. For a symmetric pitch vector with $g_x = g_y$, the spin texture propagates along the diagonal direction $(\alpha = \pi/4)$; hence, along the perpendicular direction $(\alpha = 3\pi/4)$, the system retains inversion symmetry. This directional restoration of inversion symmetry and the corresponding spectral symmetry in the $k_x = -k_y$ direction also holds for Shiba lattices with planar helical spin textures~\cite{Pritam_2024}.

Then, we investigate the BdG band gap closing at finite Cooper pair momentum, a key condition for the onset of the superconducting diode effect (SDE). For an in-plane spin texture, the gap closes at symmetric Cooper pair momenta $\mathbf{q} = \pm \mathbf{q}_1$ [see \fig{fig:band}(d)]. In contrast, the conical spin texture leads to gap closing at asymmetric values of $\mathbf{q}$ [\fig{fig:band}(e)], a direct consequence of broken inversion symmetry, which gives rise to the diode effect. Notably, along the perpendicular direction $k_x = -k_y$, the gap closing remains asymmetric and occurs at a finite Cooper pair momentum as depicted in \fig{fig:band}(f).

\section{Possibility of SDE considering different spin textures}
\subsection{Planar spin texture configuration}

\begin{figure}[t]
    \centering 
    \includegraphics[width=1\columnwidth]{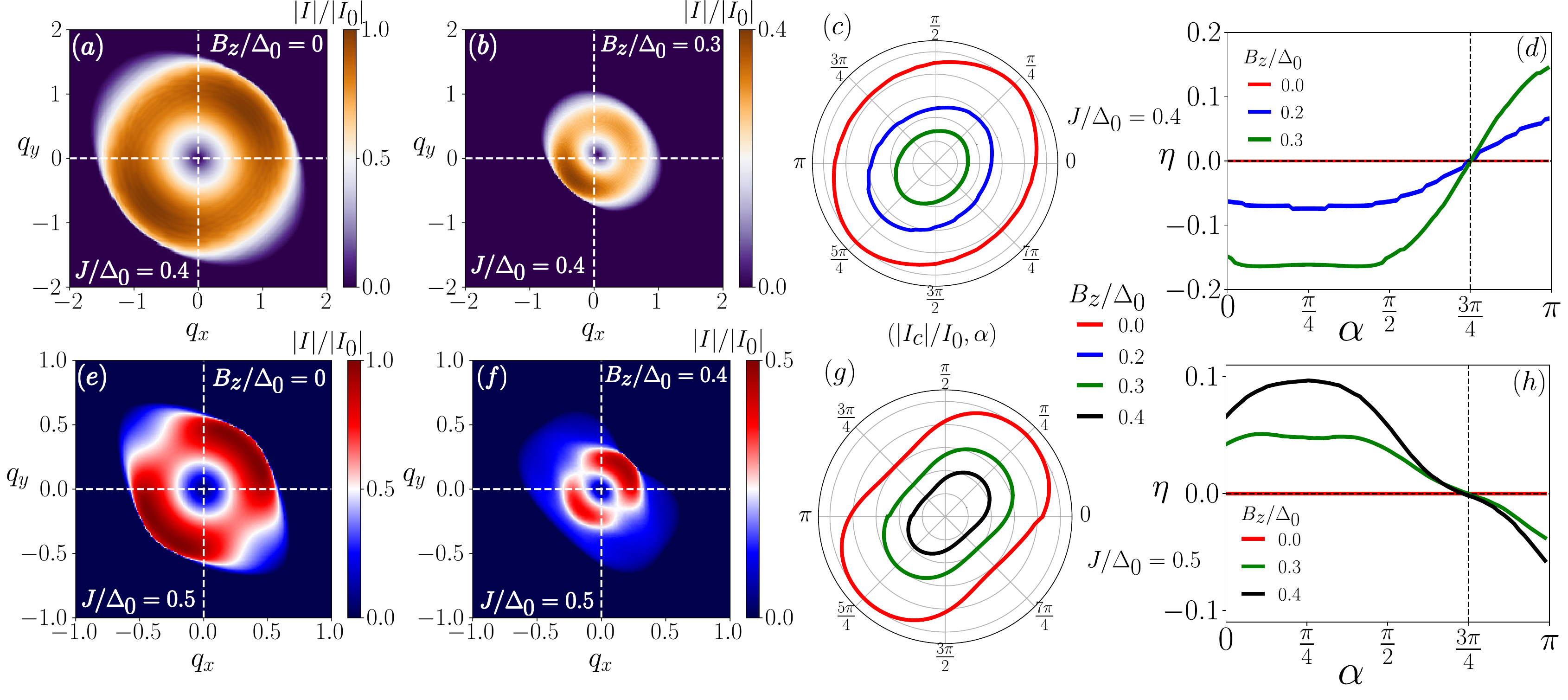}
    \caption{\textbf{Non reciprocal critical currents for planar spin texture configuration:} Panels (a)-(d) depicts the results obtained by considering the lattice Hamiltonian [Eq.~(\ref{eq:lattice2})] featuring planar spin texture. The super-current density $\vect{I/I_0}$ is shown as a function of Cooper pair momentum $\mathbf{q}$ ($q_x - q_y$ plane) for $B_z/\Delta_0=0$, $B_z/\Delta_0=0.3$ in panels (a), (b) respectively with $J/\Delta_0=0.4$. The  angular dependence $\alpha$ of critical current $\vect{I_c}(|I_c|,\alpha)$ and diode efficiency $\eta$ is shown for different values of $B_z$ in panels (c), (d) respectively. Panels (e)-(h) depicts the corresponding results obtained by considering the  effective lattice Hamiltonian  Eq.~(\ref{eq:ham_real2}).  The super-current density $\vect{I/I_0}$ is shown as a function of Cooper pair momentum $\mathbf{q}$ ($q_x - q_y$ plane) for $B_z/\Delta_0=0$, $B_z/\Delta_0=0.4$ in panels (e), (f) with $J/\Delta_0=0.5$ respectively. The  angular dependence $\alpha$ of critical current $\vect{I_c}(|I_c|,\alpha)$ and diode efficiency $\eta$ are shown for different values of $B_z$ in panels (g), (h) respectively. The BCS gap $\Delta_0(J,B_z, \mu, \mathbf{q} = 0)$ is set to $1~\mathrm{meV}$, considering a $L_x \times L_y = (25 \times 25)$ two-dimensional square lattice. Other model parameters  are considered as $(g_x,g_y,\mu/\Delta_0,t/\Delta_0,\beta^{-1}/\Delta_0)=(\pi/2,\pi/2,1,0.5,0.1)$.} 
    \label{fig:PLANAR}
\end{figure}

In this section, we discuss the potential of realizing SDE in a similar setup consisting of an in-plane helical spin texture ($\theta=\pi/2$) deposited on the surface of an $s$-wave superconducting substrate.  Employing the same theoretical framework as mentioned in the main text by considering both the lattice Hamiltonian Eq.~(\ref{eq:lattice2}) and the effective lattice regularized Hamiltonian Eq.~(\ref*{eq:ham_real2})),  we show that this setup manifests SDE only when subjected to an external out-of plane Zeeman filed $B_z$. The resulting behavior closely mimics that of the setup characterized by a conical spin texture configuration ($0 < \theta < \pi/2$) as mentioned in the main text. 

The 2D lattice Hamiltonian of the following setup consisting of  the in-plane helical spin texture and an out-of plane external Zeeman field 
can be written as 
\begin{equation}
	\begin{aligned}
		{H_l}_1=&\sum_{<i,j> ,\alpha}(t c^{\dagger}_{i,\alpha}c_{j,\alpha}+h.c.)+J\sum_{i,\alpha,\beta}(\vect{S^{\prime}}_{i}.\sigma)_{\alpha,\beta}c^{\dagger}_{i,\alpha}c_{i,\beta}+B_z \sum_{i,\alpha,\beta} (\sigma_z)_{\alpha,\beta}c^{\dagger}_{i,\alpha}c_{i,\beta}  \\
		&-\mu\sum_{i,\alpha}c^{\dagger}_{i,\alpha}c_{i,\alpha}+\sum_i(\Delta e^{i\vect{q}.\vect{r}_i}c^{\dagger}_{i,\uparrow}c^{\dagger}_{i,\downarrow}+h.c.)+\frac{A}{U}|\Delta|^2\ .
	\end{aligned}
	\label{eq:lattice2}
\end{equation}

The  local spin vector of the planar spin texture is considered as $\vect{S^\prime}(\vect{r}_i)$=$\lvert \vect{S^\prime_i} \rvert\begin{pmatrix} \cos[\phi(\vect{r_i})],\!&\! \sin[\phi(\vect{r_i})],\! &\! 0 \end{pmatrix}$, where, $\phi(\vect{r_i})=\vect{g}\!\cdot \!\vect{r}_i=\left( g_x x_i+g_y y_i \right)$, represents the spatial variation of the spins with $\vect{g}$ being the pitch of the spin texture. The rest of system parameters remain same as those defined in the main text. The effective lattice regularized Hamiltonian is derived as: 
\begin{align}
	{H^{\prime}_{l}}_1 =& - \sum_{\substack{\mathbf{n} \\ \sigma, \sigma^\prime}} \sum_{\mu = x, y} c_{\sigma^\prime, \mathbf{n}+\hat{\mu}}^\dagger \bigg[ t\, \delta_{\sigma,\sigma^\prime} + \frac{i}{2}\tilde{g}_\nu (\sigma^z)_{\sigma'\sigma} + \text{h.c.}\bigg] c_{\sigma, \mathbf{n}} \nonumber \\
	&+\sum_{\sigma, \sigma^\prime, \mathbf{n}} c_{\sigma^\prime, \mathbf{n}}^\dagger \bigg[ \left( 4t + \frac{\tilde{g}_{x}^2 + \tilde{g}_{y}^2}{2} - \mu \right)\delta_{\sigma,\sigma^\prime} 
	+ B_z (\sigma^z)_{\sigma^\prime \sigma} + J  (\sigma^x)_{\sigma^\prime \sigma} \bigg] c_{\sigma, \mathbf{n}} \nonumber \\
	& + \sum_{\mathbf{n}} \left( \Delta e^{i \mathbf{q}\cdot\mathbf{n}} c_{\mathbf{n},\ua}^\dagger c_{\mathbf{n},\da}^\dagger + h.c. \right) + \frac{L_x L_y}{U} |\Delta|^2\ . \label{eq:ham_real2}
\end{align}

\begin{figure}[b]
\centering 
\includegraphics[width=0.65\columnwidth]{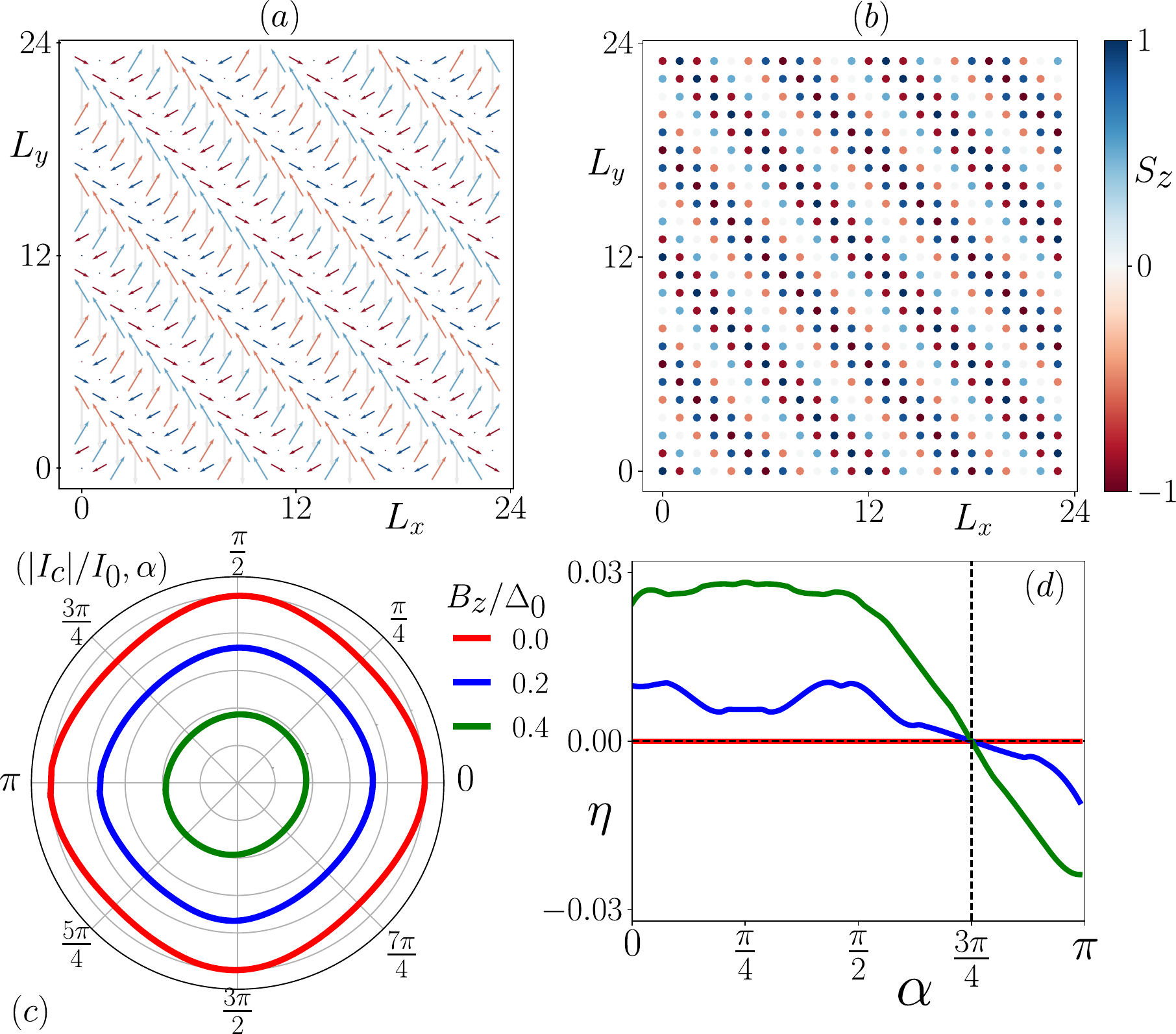}
\caption{\textbf{Non-reciprocal critical currents for AFM spin texture configuration:} The spatial profile of the AFM spin texture configuration in a $24\times24$ square lattice is demonstrated, where panel (a) depicts the in-plane components ($S_{x i}, S_{y i}$) of the local spin vectors using arrows, 
and panel (b) displays the out of plane component $S_{z i}$ with color scale indicating the magnitude and sign of $S_{z i}$. The  angular dependence $\alpha$ of critical current $\vect{I_c}(|I_c|,\alpha)$ and diode efficiency $\eta$ are shown for different values of $B_z$ in 
panels (c), (d) respectively. Here, $I_0 = I_c$ ($B_z / \Delta_0 = 0.0$, $\mu / \Delta_0 = 0)$. The other system parameters are chosen as $(g_x,g_y,h_x,h_y,\mu/\Delta_0,t/\Delta_0, \beta^{-1}/\Delta_0)=(\pi/5,\pi/5,\pi/6,\pi/6,0.5,0.5,0.1)$.}
\label{fig:AFM}
\end{figure}

Having obtained the self-consistent solution $\Delta(q)$, following the similar analysis as discussed in the main text and computing the super-current density $\vect{I}$ via Eq.~(8) of the main text, we show that the magnitude of the super-current, $|\vect{I}|$, exhibits a symmetric profile in the $q_x$–$q_y$ plane with respect to $\vect{q}$ when the out-of-plane Zeeman field is absent ($B_z = 0$) (see Fig.~\ref{fig:PLANAR}(a), (e)). In contrast, for a finite  Zeeman field $B_z\ne0$, the magnitude of the supercurrent $|\vect{I}|$ develops a clear asymmetry in ${q_x-q_y}$ plane, as evident from Fig.~\ref{fig:PLANAR}(b), (f), hence indicating non-reciprocal behavior. This asymmetry becomes especially evident along the  $q_x = q_y$ line for the symmetric choice of the pitch vector of the spin texture ($g_x=g_y$). Then, the critical current $I_c$, obtained my maximizing $|I|$ for each value of $\alpha$, is reciprocal $I_c(\alpha)=I_c(\alpha+\pi)$ for all $\alpha$ when $B_z=0$. As shown in Fig.~{\ref{fig:PLANAR}(c), (g)}, non-reciprocity in critical current is achieved resulting in a finite diode efficiency, when $B_z\ne0$.  As demonstrated in Fig.~{\ref{fig:PLANAR}(d), (h)}, the diode efficiency $\eta$ as a function of $\alpha$ displays a similar behavior to the characteristic features when the spin texture with cone phase is considered. All the qualitative features can be understood from the symmetries of the underlying spin texture, following the same arguments outlined in the main text during the discussion of the cone phase. Notably, similar behavior has also been reported in 1D Shiba chains with planar helical spin textures under external magnetic fields, where the combined effect of spin winding and Zeeman coupling gives rise to spectral asymmetry and the SDE~\cite{Sayak_2025}.

\subsection{Antiferromagnetic (AFM) spin texture configuration}

The local spin vector for the AFM spin texture at the $i^{\rm{th}}$ lattice site is defined for the lattice Hamiltonian Eq.~(\ref{eq:lattice2})  as:
 $\vect{S^{\prime}}_i$=$\lvert \vect{S^{\prime}_i} \rvert\begin{pmatrix} \sin[{\theta(\vect{r_i})}]\cos[\phi(\vect{r_i})],\!&\! \sin[{\theta(\vect{r_i})}]\sin[\phi(\vect{r_i})],\! &\! \cos[{\theta(\vect{r_i})}] \end{pmatrix}$  where, $|\mathbf{S}'_i|=1$ is considered 
 and the angular functions vary over the lattice as:
\begin{equation}
	\phi(\mathbf{r}_i) = \mathbf{g} \cdot \mathbf{r}_i = g_x x_i + g_y y_i, \quad
	\theta(\mathbf{r}_i) = (-1)^{x_i + y_i} \, \mathbf{h} \cdot \mathbf{r}_i = (-1)^{x_i + y_i} \left( h_x x_i + h_y y_i \right)\ .
	\label{eq:afm_angle}
\end{equation}

The AFM spin texture is illustrated in Fig.~\ref{fig:AFM}(a),(b) considering a $24\times24$ square lattice. The spatial variation of the planar ($x–y$ plane) components of the spin vector is depicted in Fig.~\ref{fig:AFM}(a), with the color code representing the magnitude of the out-of-plane component, ${S_z}_i$. The component ${S_z}_i$ varies spatially across the lattice within the range $(-1, 1)$ with $\langle S_z \rangle = 0$ in the setup (see Fig.~\ref{fig:AFM}(b)). Next, by performing a similar self-consistent analysis and computing the critical currents considering the AFM spin texture configuration, we demonstrate that the system exhibits non-reciprocal critical current behavior leading to a finite diode efficieny $\eta$, only in the presence of an out-of-plane Zeeman field $B_z$ (see Figs.~\ref{fig:AFM}(c),(d)). The efficiency $\eta$ as a function of $\alpha$ displays qualitatively similar trends for the symmetric choice $g_x = g_y$, as observed in setups with helical or conical spin texture. However, in contrast, implementation of the AFM spin texture yields significantly low diode efficiencies ($\eta \sim 3\%$) (see Figs.~\ref{fig:AFM}(c),(d)). Notably, while the overall qualitative behavior remains similar, the diode efficiency for the AFM spin texture setup can be further optimized by tuning various system parameters.

\end{document}